\documentclass[12pt]{article}
\usepackage{amsmath,amssymb}
\usepackage{bm}
\usepackage{mathrsfs}
\usepackage{fourier}
\usepackage{multirow}
\allowdisplaybreaks[4]

\usepackage[usenames]{color}

\begin{document}

\title{Orbifold Family Unification on 6 Dimensions}

\author{
Yuhei \textsc{Goto}\footnote{E-mail: 12SM207K@shinshu-u.ac.jp},~~~
Yoshiharu \textsc{Kawamura}\footnote{E-mail: haru@azusa.shinshu-u.ac.jp}\\
{\it Department of Physics, Shinshu University, }\\
{\it Matsumoto 390-8621, Japan}\\
and \\
Takashi \textsc{Miura}\footnote{E-mail: takashi.miura@people.kobe-u.ac.jp}\\
{\it Department of Physics, Kobe University, }\\
{\it Kobe 657-8501, Japan}\\
}

\date{
July 10, 2013}

\maketitle
\begin{abstract}
We study the possibility of family unification on the basis of $SU(N)$ gauge theory
on the 6-dimensional space-time, $M^4\times T^2/Z_N$.
We obtain enormous numbers of models with three families of $SU(5)$ matter multiplets 
and those with three families of the standard model multiplets,
from a single massless Dirac fermion with a higher-dimensional representation of $SU(N)$, 
through the orbifold breaking mechanism.
\end{abstract}


\section{Introduction}

The origin of the family replication has been a big riddle.
The {\it family unification} based on a large symmetry group can provide a possible solution.
The studies have been carried out intensively,
and they are classified into two categories. 
One is the investigation based on the 4-dimensional Minkowski space-time~\cite{R,G,F1,F2,K&Y},
and the other is that 
based on higher-dimensional space-times~\cite{BB&K,Watari:2002tf,GMN1,GLMN,GLMS,GMN2,KK&O,K&M,FKMN&S}.

The advantage of higher-dimensional theories is 
that substances including mirror particles can be reduced using the symmetry breaking mechanism 
concerning extra dimensions, as originally discussed in superstring theory~\cite{CHS&W,DHV&W1,DHV&W2}.
Here, the mirror particles are particles with opposite quantum numbers under the standard model (SM) gauge group.
Hence, a candidate realizing the family unification
is grand unified theories (GUTs) on a higher-dimensional space-time
including an orbifold as an extra space.\footnote{
5-dimensional supersymmetric GUTs on $M^4 \times S^1/Z_2$ possess the attractive feature that
the triplet-doublet splitting of Higgs multiplets is elegantly realized~\cite{K,H&N}.} 

In this paper, we study the possibility of family unification on the basis of $SU(N)$ gauge theory
on $M^4\times T^2/Z_N$, using the method in Ref.~\cite{KK&O}.
We investigate whether or not three families are derived from a single massless Dirac fermion of $SU(N)$
for two patterns of symmetry breaking.

The contents of this paper are as follows.
In Sec. II, we provide general arguments on the orbifold breaking based on 2-dimensional orbifold $T^2/Z_N$
and formulae for numbers of species.
In Sec. III, we investigate the family unification for each $T^2/Z_N$ $(N=2, 3, 4, 6)$,
in the framework of 6-dimensional $SU(N)$ GUTs. 
Section IV is devoted to conclusions and discussions.

\section{$Z_N$ orbifold breaking and formulae for numbers of species}

We explain the orbifold $T^2/Z_N$
and give formulae for numbers of species, 
in the case with diagonal embeddings for representation matrices of $Z_N$ transformations.

\subsection{$Z_N$ orbifold breaking}

Let $z$ be the complex coordinate of $T^2/Z_N$.
Here, $T^2$ is constructed from a 2-dimensional lattice.
On $T^2$, the points $z + e_1$ and $z + e_2$ are identified with
the point $z$, where $e_1$ and $e_2$ are basis vectors.
The orbifold $T^2/Z_N$ is obtained by dividing $T^2$ by the $Z_N$ transformation
$Z_N:z \to \xi z$ $(\xi^N = 1)$ so that the point $z$ is identified with $\xi z$, 
or $z$ is generally identified with
$\xi^k z + a e_1 + b e_2$, where $k$, $a$ and $b$ are integers.

Let us explain the orbifold breaking using $T^2/Z_2$.
Accompanied by the identification of points on $T^2/Z_2$,
the following boundary conditions for a field $\Phi(x, z)$ can be imposed on,
\begin{eqnarray}
&~& \Phi(x, -z) =T_{\Phi}[P_0] \Phi(x,z)~,~~ \Phi(x, e_1-z) =T_{\Phi}[P_1] \Phi(x,z)~,~~
\nonumber \\ 
&~& \Phi(x, e_2-z) =T_{\Phi}[P_2] \Phi(x,z)~,
\label{T[P]}
\end{eqnarray}
where $e_1 =1$, $e_2 =i$, and
$T_{\Phi}[P_0]$, $T_{\Phi}[P_1]$ and $T_{\Phi}[P_2]$ represent appropriate representation matrices.
The $P_0$, $P_1$ and $P_2$ stand for the representation matrices of 
the $Z_2$ transformations $z \to -z$, $z \to e_1-z$ and $z \to e_2-z$
for fields with the fundamental representation.

The eigenvalues of $T_{\Phi}[P_0]$, $T_{\Phi}[P_1]$  and $T_{\Phi}[P_2]$ 
are interpreted as the $Z_2$ parities for the extra space.
The fields with even $Z_2$ parities have zero modes, 
but those including an odd $Z_2$ parity do not have zero modes.
Here, zero modes mean 4-dimensional massless fields surviving after compactification.
Kaluza-Klein modes do not appear in our low-energy world,
because they have heavy masses of $O(1/R)$, with the same magnitude 
as the unification scale.
{\it Unless all components of non-singlet field have a common $Z_2$ parity, 
a symmetry reduction occurs upon compactification because zero modes are absent
in fields with an odd parity.}
This type of symmetry breaking mechanism is called $\lq\lq$orbifold breaking mechanism''.\footnote{
The $Z_2$ orbifolding was used in superstring theory~\cite{A} and heterotic $M$-theory~\cite{H&W1,H&W2}.
In field theoretical models, it was applied to the reduction of global SUSY~\cite{M&P,P&Q}, which is
an orbifold version of Scherk-Schwarz mechanism~\cite{S&S,S&S2}, and then
to the reduction of gauge symmetry~\cite{K1}.
}

Basis vectors, representation matrices and their transformation properties of $T^2/Z_N$
are summarized in Table \ref{T1}~\cite{K&M3,K&M2}.\footnote{
Though the number of independent representation matrices for $T^2/Z_6$ is stated to be three in \cite{K&M},
it should be two because other operations are generated using $s_0:z \to e^{\pi i/3} z$ and $r_1:z \to e_1-z$.
For example, $t_1:z \to z+e_1$ and $t_2:z \to z+e_2$ are generated as $t_1= r_1 (s_0)^3$
and $t_2 = (s_0)^2 r_1 (s_0)^4 r_1$, respectively.
}
\begin{table}[htb]
\caption{The characters of $T^2/Z_N$.}
\label{T1}
\begin{center}
\begin{tabular}{c|c|c|c} \hline
$N$ & {\it Basis vectors} & {\it Rep. matrices} & {\it Transformation properties} \\ \hline\hline
$2$ & $1, i$ & $P_0,~ P_1,~ P_2$ & $z \to -z,~ z \to e_1-z,~ z \to e_2-z$ \\ \hline
$3$ & $1, e^{2\pi i/3}$ & $\Theta_0,~ \Theta_1$ & $z \to e^{2\pi i/3} z,~ z \to e^{2\pi i/3} z + e_1$ \\ \hline
$4$ & $1, i$ & $Q_0,~ P_1$ & $z \to iz,~ z \to e_1-z$ \\ \hline
$6$ & $1, (-3+i\sqrt{3})/2$ & $\Xi_0,~ P_{1}$ & $z \to e^{\pi i/3} z,~ z \to e_1-z$ \\ \hline
\end{tabular}
\end{center}
\end{table}
Note that there is a choice in representation matrices, and
$P_1$ concerning the $Z_2$ transformation $z \to e_1-z$ is also used in $T^2/Z_4$ and $T^2/Z_6$.

Fields possess discrete charges relating eigenvalues of representation matrices
for $Z_M$ transformation.
Here, $M=N$ for $N=2,3$ and $M=N,2$ for $N=4,6$.
The discrete charges are assigned as numbers $n/M$ $(n=0, 1, \cdots, M-1)$
and $e^{2\pi i n/M}$ are elements of $Z_M$ transformation.
We refer to them as $Z_M$ elements.

A fermion with spin $1/2$ in 6-dimensions is regarded as a Dirac fermion 
or a pair of Weyl fermions with opposite chiralities in 4-dimensions.
There are two choices in a 6-dimensional Weyl fermion, i.e.,
\begin{eqnarray}
&~& \Psi_+ = \frac{1+\Gamma_7}{2} \Psi 
= \left(
\begin{array}{cc}
\frac{1-\gamma_5}{2} & 0 \\
0 & \frac{1+\gamma_5}{2} 
\end{array}
\right)
\left( 
\begin{array}{c}
\Psi^1 \\
\Psi^2 
\end{array}
\right)
= \left( 
\begin{array}{c}
\Psi^1_L \\
\Psi^2_R 
\end{array}
\right)~,
\label{Psi+}\\
&~& \Psi_- = \frac{1-\Gamma_7}{2} \Psi
= \left(
\begin{array}{cc}
\frac{1+\gamma_5}{2} & 0 \\
0 & \frac{1-\gamma_5}{2} 
\end{array}
\right)
\left( 
\begin{array}{c}
\Psi^1 \\
\Psi^2 
\end{array}
\right)
= \left( 
\begin{array}{c}
\Psi^1_R \\
\Psi^2_L 
\end{array}
\right)~,
\label{Psi-}
\end{eqnarray}
where $\Psi_+$ and $\Psi_-$ are fermions with positive
and negative chirality, respectively, and
$\Gamma_7$ and $\gamma_5$ are the chirality operators for 6-dimensional fermions
and 4-dimensional ones, respectively.\footnote{
For more detailed explanations 
for 6-dimensional fermions, see Ref.~\cite{K&M4}.
}
Here and hereafter, the subscript $\pm$ stands for the chiralities on 6 dimensions.

From the $Z_M$ invariance of kinetic term
and the transformation property of the covariant derivatives 
$Z_M : D_z \to \overline{\rho} D_z$ and $D_{\overline{z}} \to \rho D_{\overline{z}}$
with $\overline{\rho} = e^{-2\pi i/M}$ and $\rho = e^{2\pi i/M}$,
the following relations hold between
the $Z_M$ element of $\Psi^1_{L(R)}$ and $\Psi^2_{R(L)}$,
\begin{eqnarray}
\mathcal{P}_{\Psi^2_R} = {\rho} {\mathcal{P}}_{\Psi^1_{L}}~,~~
\mathcal{P}_{\Psi^1_R} = \overline{\rho} {\mathcal{P}}_{\Psi^2_{L}}~,
\label{rhoPsiR}
\end{eqnarray}
where $z \equiv x^5 + i x^6$ and $\overline{z} \equiv x^5 - i x^6$.

Chiral gauge theories including Weyl fermions on even dimensional space-time
become, in general, anomalous in the presence of gauge anomalies,
gravitational anomalies, mixed anomalies and/or global anomaly~\cite{D&P,BG&T}.
In $SU(N)$ GUTs on 6-dimensional space-time,
the global anomaly is absent because of $\Pi_6(SU(N)) = 0$ for $N \ge 4$.
Here, $\Pi_6(SU(N))$ is the 6-th homotopy group of $SU(N)$.
In our analysis, we consider a massless Dirac fermion $(\Psi_+, \Psi_-)$ under the $SU(N)$ gauge group $(N \ge 8)$
on 6-dimensional space-time.
In this case, anomalies are canceled out by the contributions from fermions with different chiralities
\subsection{Formulae for numbers of species}

With suitable diagonal representation matrices $R_a$ 
($a=0, 1, 2$ for $T^2/Z_2$ and $a=0, 1$ for $T^2/Z_3$, $T^2/Z_4$ and $T^2/Z_6$),
the $SU(N)$ gauge group is broken down into its subgroup such that
\begin{eqnarray}
SU(N) \to SU(p_1)\times SU(p_2) \times \cdots \times SU(p_{n})\times U(1)^{n-m-1}~,
\label{GSB}
\end{eqnarray}
where $N=\sum_{i=1}^{n} p_i$.
Here and hereafter, $SU(1)$ unconventionally stands for $U(1)$, $SU(0)$ means nothing
and $m$ is a sum of the number of $SU(0)$ and $SU(1)$. 
The concrete form of $R_a$ will be given in the next section.

After the breakdown of $SU(N)$, the rank $k$ totally antisymmetric tensor representation $[N, k]$, 
whose dimension is ${}_{N}C_{k}$,
is decomposed into a sum of multiplets of the subgroup 
$SU(p_1) \times \cdots \times SU(p_n)$ as
\begin{eqnarray}
[N, k] = \sum_{l_1 =0}^{k} \sum_{l_2 = 0}^{k-l_1} \cdots \sum_{l_{n-1} = 0}^{k-l_1-\cdots -l_{n-2}}  
\left({}_{p_1}C_{l_1}, {}_{p_2}C_{l_2}, \cdots, {}_{p_n}C_{l_n}\right) ,
\label{Nk}
\end{eqnarray}
where $l_n=k-l_1- \cdots -l_{n-1}$ and our notation is that ${}_{n}C_{l} = 0$ for $l > n$ and $l < 0$.
Here and hereafter, we use ${}_{n}C_{l}$ instead of $[n, l]$ in many cases.
We sometimes use the ordinary notation for representations too, 
e.g., ${\bf{5}}$ and ${\overline{\bf{5}}}$ in place of ${}_{5}C_{1}$ and ${}_{5}C_{4}$. 

The $[N, k]$ is constructed by the antisymmetrization of $k$-ple product of 
the fundamental representation ${\bm{N}} = [N, 1]$:
\begin{eqnarray}
[N, k] = ({\bm{N}} \times \dots \times {\bm{N}})_{\tiny {\mbox{A}}}~.
\label{N*...*N} 
\end{eqnarray}
We define the intrinsic $Z_M$ elements $\eta^a_k$ such that
\begin{eqnarray}
({\bm{N}} \times \dots \times {\bm{N}})_{\tiny {\mbox{A}}} 
\to \eta^a_k (R_a {\bm{N}} \times \dots \times R_a {\bm{N}})_{\tiny {\mbox{A}}}~.
\label{etaNka}
\end{eqnarray}
By definition, $\eta^a_k$ take a value of $Z_M$ elements, i.e.,
$e^{2\pi i n/M}$ $(n=0, 1, \cdots, M-1)$.
Note that $\eta^a_k$ for $\Psi_+$ are not necessarily same as those of $\Psi_-$,
and the chiral symmetry is still respected.

Let us investigate the family unification in two cases.
Each breaking pattern is given by
\begin{eqnarray}
&~& SU(N) \to  SU(5) \times SU(p_2) \times \cdots \times SU(p_n) \times U(1)^{n-m-1}~,
\label{OB1}\\
&~& SU(N) \to  SU(3) \times SU(2) \times SU(p_3) \times \cdots \times SU(p_n) \times U(1)^{n-m-1}~,
\label{OB2}
\end{eqnarray}
where $SU(3)$ and $SU(2)$ are identified with $SU(3)_C$ and $SU(2)_L$ in 
the SM gauge group.

\subsubsection{Formulae for $SU(5)$ multiplets}

We study the breaking pattern (\ref{OB1}).
After the breakdown of $SU(N)$, $[N, k]$ is decomposed as
\begin{eqnarray}
[N, k] = \sum_{l_1 =0}^{k} \sum_{l_2 = 0}^{k-l_1} \cdots \sum_{l_{n-1} = 0}^{k-l_1-\cdots -l_{n-2}}  
\left({}_{5}C_{l_1}, {}_{p_2}C_{l_2}, \cdots, {}_{p_n}C_{l_n}\right)~.
\label{Nk(p1=5)}
\end{eqnarray}
As mentioned before, 
 ${{}_{5}C_{0}}$, ${{}_{5}C_{1}}$, ${{}_{5}C_{2}}$, ${{}_{5}C_{3}}$, ${{}_{5}C_{4}}$
and ${{}_{5}C_{5}}$ stand for representations ${\bf{1}}$, ${\bf{5}}$, ${\bf{10}}$, 
 ${\overline{\bf{10}}}$, ${\overline{\bf{5}}}$ and ${\overline{\bf{1}}}$.\footnote{
We denote the $SU(5)$ singlet relating to ${{}_{5}C_{5}}$ as ${\overline{\bf{1}}}$, for convenience sake,
to avoid the confusion over singlets.}

Utilizing $\lq\lq$survival hypothesis'' and the equivalence of $({\bf{5}}_R)^c$ and $(\overline{\bf{10}}_R)^c$
with $\overline{\bf{5}}_L$ and ${\bf{10}}_L$, respectively,
\footnote{
As usual, $({\bf{5}}_R)^c$ and $(\overline{\bf{10}}_R)^c$ represent 
the charge conjugate of ${\bf{5}}_R$ and $\overline{\bf{10}}_R$, respectively. 
Note that $({\bf{5}}_R)^c$ and $(\overline{\bf{10}}_R)^c$ transform as the left-handed Weyl fermions 
under the 4-dimensional Lorentz transformations.
}
we write the numbers of $\overline{\bf 5}$ and ${\bf{10}}$ representations 
for left-handed Weyl fermions as
\begin{eqnarray}
&~& n_{\bar{5}} \equiv \sharp{\overline{\bf 5}}_L  - \sharp{\bf 5}_L + \sharp{\bf 5}_R  - \sharp{\overline{\bf 5}}_R~,  
\label{nbar5-def}\\
&~& n_{10} \equiv \sharp{\bf 10}_L  - \sharp{\overline{\bf 10}}_L + \sharp{\overline{\bf 10}}_R  - \sharp{\bf 10}_R~, 
\label{n10-def}
\end{eqnarray}
where $\sharp$ represents the number of each multiplet. 
Here, the survival hypothesis is the assumption that 
{\it if a symmetry is broken down into a smaller symmetry at a scale $M_{\tiny{\mbox{SB}}}$,
then any fermion mass terms invariant under the smaller group induce fermion masses 
of order $O(M_{\tiny{\mbox{SB}}})$}~\cite{G,BNM&S}.

The $SU(5)$ singlets are regarded as the right-handed neutrinos, 
which can obtain heavy Majorana masses among themselves as well as the Dirac masses with left-handed neutrinos.
Some of them can be involved in see-saw mechanism~\cite{ss1,ss2,ss3}.
The total number of $SU(5)$ singlets (with heavy masses) is given by
\begin{eqnarray}
n_{1} \equiv \sharp{\bf 1}_L  + \sharp{\overline{\bf 1}}_L + \sharp{\overline{\bf 1}}_R  + \sharp{\bf 1}_R~.
\label{n1-def}
\end{eqnarray}

Formulae for $n_{\bar{5}}$, $n_{10}$ and $n_{1}$ from
a Dirac fermion $(\Psi_+, \Psi_-)$ whose intrinsic $Z_M$ elements are $(\eta_{k+}^a, \eta_{k-}^a)$
are given by
\begin{eqnarray}
&~& n_{\bar{5}} = \sum_{\pm} \sum_{l_1 = 1, 4} (-1)^{l_1}
\left(\sum_{\{l_2, \cdots, l_{n-1}\}_{n^a_{l_1 L \pm}}} - \sum_{\{l_2, \cdots, l_{n-1}\}_{n^a_{l_1 R \pm}}}\right)
{}_{p_2}C_{l_2} \cdots {}_{p_{n}}C_{l_{n}}~, 
\label{nbar5-ZM}\\
&~& n_{10} = \sum_{\pm} \sum_{l_1 = 2, 3} (-1)^{l_1}
\left(\sum_{\{l_2, \cdots, l_{n-1}\}_{n^a_{l_1 L \pm}}} - \sum_{\{l_2, \cdots, l_{n-1}\}_{n^a_{l_1 R \pm}}}\right)
{}_{p_2}C_{l_2} \cdots {}_{p_{n}}C_{l_{n}}~, 
\label{n10-ZM}\\
&~& n_{1} = \sum_{\pm} \sum_{l_1 = 0, 5}
\left(\sum_{\{l_2, \cdots, l_{n-1}\}_{n^a_{l_1 L \pm}}} + \sum_{\{l_2, \cdots, l_{n-1}\}_{n^a_{l_1 R \pm}}}\right)
{}_{p_2}C_{l_2} \cdots {}_{p_{n}}C_{l_{n}}~,
\label{n1-ZM}
\end{eqnarray}
where $p_{n} = N-\sum_{i=1}^{n-1} p_i$ and $l_{n} = N-\sum_{i=1}^{n-1} l_i$.
$\sum_{\pm}$ represents the summation of contributions from $\Psi_+$ and $\Psi_-$.
Furthermore, $\sum_{\{l_2, \cdots, l_{n-1}\}_{n^a_{l_1 L \pm}}}$ means that
the summations over $l_j=0, \cdots, k-l_1-\cdots - l_{j-1}$ $(j=2, \cdots, n-1)$
are carried out under the condition that $l_j$ should satisfy
specific relations on $T^2/Z_N$ given in Table \ref{T2}.
\begin{table}[htb]
\caption{The specific relations for $l_j$.}
\label{T2}
\begin{center}
\begin{tabular}{c||l|l} \hline
{\it Orbifolds} & $\overline{\rho}^{k}\eta^a_{k \pm}$ & {\it Specific relations}  \\ \hline\hline
$T^2/Z_2$ & $(-1)^{k}\eta^0_{k \pm}=(-1)^{\alpha_\pm}$ 
& $n^0_{l_1 L \pm} \equiv l_2+l_3+l_4 = 2-l_1 - \alpha_{\pm}~~~(\mbox{mod}~2)$ \\
& $(-1)^{k}\eta^1_{k \pm}=(-1)^{\beta_{\pm}}$ 
& $n^1_{l_1 L \pm} \equiv l_2+l_5+l_6 = 2-l_1 - \beta_{\pm}~~~(\mbox{mod}~2)$ \\
& $(-1)^{k}\eta^2_{k \pm} = (-1)^{\gamma_{\pm}}$ 
& $n^2_{l_1 L \pm} \equiv l_3+l_5+l_7 = 2 - l_1 - \gamma_{\pm} ~~~(\mbox{mod}~2)$ \\ \hline
$T^2/Z_3$ & $(e^{-2\pi i/3})^{k}\eta^0_{k \pm} =(e^{2\pi i/3})^{\alpha_{\pm}}$ & 
$n^0_{l_1 L \pm} \equiv l_2+l_3+2(l_4+l_5+l_6)$\\
& &~~~~~~~~~~~$= 3-l_1 - \alpha_{\pm}~~~(\mbox{mod}~3)$\\
& $(e^{-2\pi i/3})^{k}\eta^1_{k \pm}=(e^{2\pi i/3})^{\beta_{\pm}}$ & 
$n^1_{l_1 L \pm} \equiv l_4+l_7+2(l_2+l_5+l_8)$\\
& &~~~~~~~~~~~$= 3- l_1 - \beta_{\pm}~~~(\mbox{mod}~3)$ \\ \hline
$T^2/Z_4$ & $(-i)^{k}\eta^0_{k \pm} =i^{\alpha_{\pm}}$ & 
$n^0_{l_1 L \pm} \equiv l_2+2(l_3+l_4)+3(l_5+l_6)$ \\
& &~~~~~~~~~~~$= 4-l_1 - \alpha_{\pm}~~~(\mbox{mod}~4)$ \\
& $(-1)^{k}\eta^1_{k \pm}=(-1)^{\beta_{\pm}}$ & 
$n^1_{l_1 L \pm} \equiv l_3+l_5 + l_7 = 2- l_1 - \beta_{\pm}~~~(\mbox{mod}~2)$ \\ \hline
$T^2/Z_6$ & $(e^{-\pi i/3})^{k}\eta^0_{k \pm} =(e^{\pi i/3})^{\alpha_{\pm}}$ & 
$n^0_{l_1 L \pm} \equiv l_2+2(l_3+l_4)+3(l_5+l_6)$\\
& &~~~~~~~~~~~~~~~$+ 4(l_7+l_8) + 5(l_9+l_{10})$\\
& &~~~~~~~~~~~$=6-l_1 - \alpha_{\pm}~~~(\mbox{mod}~6)$ \\
& $(-1)^{k}\eta^1_{k \pm}=(-1)^{\beta_{\pm}}$ & 
$n^1_{l_1 L \pm} \equiv l_3+l_5 + l_7 + l_9 + l_{11}$\\
& &~~~~~~~~~~~$= 2- l_1 - \beta_{\pm}~~~(\mbox{mod}~2)$ \\ \hline
\end{tabular}
\end{center}
\end{table}
The relations will be confirmed in the next section.
In the same way, $\sum_{\{l_2, \cdots, l_{n-1}\}_{n^a_{l_1 R \pm}}}$ means that
the summations over $l_j=0, \cdots, k-l_1-\cdots - l_{j-1}$ $(j=2, \cdots, n-1)$
are carried out under the condition that $l_j$ should satisfy
specific relations $n^a_{l_1 R \pm}=n^a_{l_1 L \pm} \mp 1$ (mod $M$) for $\Psi_{\pm}$. 
The formulae (\ref{nbar5-ZM}) -- (\ref{n1-ZM}) will be rewritten in more concrete form
for each $T^2/Z_N$ $(N=2, 3, 4, 6)$, by the use of projection operators, in the next section.

\subsubsection{Formulae for the SM multiplets}

We study the breaking pattern (\ref{OB2}).
After the breakdown of $SU(N)$, $[N, k]$ is decomposed as
\begin{eqnarray}
[N, k] = \sum_{l_1 =0}^{k} \sum_{l_2 = 0}^{k-l_1} \sum_{l_3 = 0}^{k-l_1-l_2} 
\cdots \sum_{l_{n-1} = 0}^{k-l_1-\cdots -l_{n-2}}  
\left({}_{3}C_{l_1}, {}_{2}C_{l_2}, {}_{p_3}C_{l_3}, \cdots, {}_{p_n}C_{l_n}\right)~.
\label{Nk(p1=3)}
\end{eqnarray}

The flavor numbers of down-type anti-quark singlets $(d_{R})^c$, 
lepton doublets $l_{L}$, up-type anti-quark singlets $(u_{R})^c$, 
positron-type lepton singlets $(e_{R})^c$, 
and quark doublets $q_{L}$ are denoted as 
$n_{\bar{d}}$, $n_l$, $n_{\bar{u}}$, $n_{\bar{e}}$ and $n_q$.
Using the survival hypothesis and the equivalence on charge conjugation, 
we define the flavor number of each chiral fermion as
\begin{eqnarray}
&~& n_{\bar{d}} \equiv \sharp ({}_{3}C_{2}, {}_{2}C_{2})_L - \sharp  ({}_{3}C_{1}, {}_{2}C_{0})_L 
+ \sharp  ({}_{3}C_{1}, {}_{2}C_{0})_R - \sharp  ({}_{3}C_{2}, {}_{2}C_{2})_R~,  
\label{nd-def}\\
&~& n_{l} \equiv \sharp  ({}_{3}C_{3}, {}_{2}C_{1})_L  - \sharp  ({}_{3}C_{0}, {}_{2}C_{1})_L 
+ \sharp  ({}_{3}C_{0}, {}_{2}C_{1})_R  - \sharp  ({}_{3}C_{3}, {}_{2}C_{1})_R~,  
\label{nl-def}\\
&~& n_{\bar{u}} \equiv \sharp  ({}_{3}C_{2}, {}_{2}C_{0})_L  - \sharp  ({}_{3}C_{1}, {}_{2}C_{2})_L 
+ \sharp  ({}_{3}C_{1}, {}_{2}C_{2})_R  - \sharp  ({}_{3}C_{2}, {}_{2}C_{0})_R~,  
\label{nu-def}\\
&~& n_{\bar{e}} \equiv \sharp  ({}_{3}C_{0}, {}_{2}C_{2})_L  - \sharp  ({}_{3}C_{3}, {}_{2}C_{0})_L 
+ \sharp  ({}_{3}C_{3}, {}_{2}C_{0})_R  - \sharp  ({}_{3}C_{0}, {}_{2}C_{2})_R~,  
\label{ne-def}\\
&~& n_{q} \equiv \sharp  ({}_{3}C_{1}, {}_{2}C_{1})_L  - \sharp  ({}_{3}C_{2}, {}_{2}C_{1})_L 
+ \sharp  ({}_{3}C_{2}, {}_{2}C_{1})_R  - \sharp  ({}_{3}C_{1}, {}_{2}C_{1})_R~,  
\label{nq-def}
\end{eqnarray}
where $\sharp$ again represents the number of each multiplet.
The total number of (heavy) neutrino singlets $(\nu_{R})^c$ is denoted $n_{\bar{\nu}}$ and defined as
\begin{eqnarray}
n_{\bar{\nu}} \equiv \sharp  ({}_{3}C_{0}, {}_{2}C_{0})_L  + \sharp  ({}_{3}C_{3}, {}_{2}C_{2})_L 
+ \sharp  ({}_{3}C_{3}, {}_{2}C_{2})_R  + \sharp  ({}_{3}C_{0}, {}_{2}C_{0})_R .  
\label{nnu-def}
\end{eqnarray}

Formulae for the SM species including neutrino singlets are given by
\begin{eqnarray}
\hspace{-1.5cm}&~& n_{\bar{d}} = \sum_{\pm} \sum_{(l_1, l_2) = (2,2),(1,0)} 
\!\!\!\!\!\!\!\!\!\!\! (-1)^{l_1+l_2}
\left(\sum_{\{l_3, \cdots, l_{n-1}\}_{n^a_{l_1 l_2 L \pm}}} - \sum_{\{l_3, \cdots, l_{n-1}\}_{n^a_{l_1 l_2 R \pm}}}\right)
{}_{p_3}C_{l_3} \cdots {}_{p_{n}}C_{l_{n}}~, 
\label{nd-ZM}\\
\hspace{-1.5cm}&~& n_{l} = \sum_{\pm} \sum_{(l_1, l_2) = (3,1),(0,1)} 
\!\!\!\!\!\!\!\!\!\!\! (-1)^{l_1+l_2}
\left(\sum_{\{l_3, \cdots, l_{n-1}\}_{n^a_{l_1 l_2 L \pm}}} - \sum_{\{l_3, \cdots, l_{n-1}\}_{n^a_{l_1 l_2 R \pm}}}\right)
{}_{p_3}C_{l_3} \cdots {}_{p_{n}}C_{l_{n}}~,  
\label{nl-ZM}\\
\hspace{-1.5cm}&~& n_{\bar{u}} = \sum_{\pm} \sum_{(l_1, l_2) = (2,0),(1,2)} 
\!\!\!\!\!\!\!\!\!\!\! (-1)^{l_1+l_2}
\left(\sum_{\{l_3, \cdots, l_{n-1}\}_{n^a_{l_1 l_2 L \pm}}} - \sum_{\{l_3, \cdots, l_{n-1}\}_{n^a_{l_1 l_2 R \pm}}}\right)
{}_{p_3}C_{l_3} \cdots {}_{p_{n}}C_{l_{n}}~,
\label{nu-ZM}\\
\hspace{-1.5cm}&~& n_{\bar{e}} = \sum_{\pm} \sum_{(l_1, l_2) = (0,2),(3,0)} 
\!\!\!\!\!\!\!\!\!\!\! (-1)^{l_1+l_2}
\left(\sum_{\{l_3, \cdots, l_{n-1}\}_{n^a_{l_1 l_2 L \pm}}} - \sum_{\{l_3, \cdots, l_{n-1}\}_{n^a_{l_1 l_2 R \pm}}}\right)
{}_{p_3}C_{l_3} \cdots {}_{p_{n}}C_{l_{n}}~, 
\label{ne-ZM}\\
\hspace{-1.5cm}&~& n_{q} = \sum_{\pm} \sum_{(l_1, l_2) = (1,1),(2,1)} 
\!\!\!\!\!\!\!\!\!\!\! (-1)^{l_1+l_2}
\left(\sum_{\{l_3, \cdots, l_{n-1}\}_{n^a_{l_1 l_2 L \pm}}} - \sum_{\{l_3, \cdots, l_{n-1}\}_{n^a_{l_1 l_2 R \pm}}}\right)
{}_{p_3}C_{l_3} \cdots {}_{p_{n}}C_{l_{n}}~, 
\label{nq-ZM}\\
\hspace{-1.5cm}&~& n_{\bar{\nu}} = \sum_{\pm} \sum_{(l_1, l_2) = (0,0),(3,2)}
\left(\sum_{\{l_3, \cdots, l_{n-1}\}_{n^a_{l_1 l_2 L \pm}}} + \sum_{\{l_3, \cdots, l_{n-1}\}_{n^a_{l_1 l_2 R \pm}}}\right)
{}_{p_3}C_{l_3} \cdots {}_{p_{n}}C_{l_{n}}~,
\label{nnu-ZM}
\end{eqnarray}
where $\sum_{\{l_3, \cdots, l_{n-1}\}_{n^a_{l_1 l_2 L \pm}}}$ means that
the summations over $l_j=0, \cdots, k-l_1-\cdots - l_{j-1}$ $(j=3, \cdots, n-1)$
are carried out under the condition that $l_j$ should satisfy
specific relations on $T^2/Z_N$ given in Table \ref{T3}.
\begin{table}[htb]
\caption{The specific relations for $l_j$.}
\label{T3}
\begin{center}
\begin{tabular}{c||l|l} \hline
{\it Orbifolds} & $\overline{\rho}^{k}\eta^a_{k \pm}$ & {\it Specific relations}  \\ \hline\hline
$T^2/Z_2$ & $(-1)^{k}\eta^0_{k \pm}=(-1)^{\alpha_{\pm}}$ 
& $n^0_{l_1 l_2 L \pm} \equiv l_3+l_4 = 2-l_1 -l_2- \alpha_{\pm}~~~(\mbox{mod}~2)$ \\
& $(-1)^{k}\eta^1_{k \pm}=(-1)^{\beta_{\pm}}$ 
& $n^1_{l_1 l_2 L \pm} \equiv l_5+l_6 = 2-l_1-l_2 - \beta_{\pm}~~~(\mbox{mod}~2)$ \\
& $(-1)^{k}\eta^2_{k \pm} = (-1)^{\gamma_{\pm}}$ 
& $n^2_{l_1 l_2 L \pm} \equiv l_3+l_5+l_7 = 2 - l_1 - \gamma_{\pm} ~~~(\mbox{mod}~2)$ \\ \hline
$T^2/Z_3$ & $(e^{-2\pi i/3})^{k}\eta^0_{k \pm}=(e^{2\pi i/3})^{\alpha_{\pm}}$ & 
$n^0_{l_1 l_2 L \pm} \equiv l_3+2(l_4+l_5+l_6)$\\
& &~~~~~~~~~~~~~~$= 3-l_1 -l_2- \alpha_{\pm}~~~(\mbox{mod}~3)$\\
& $(e^{-2\pi i/3})^{k}\eta^1_{k \pm} =(e^{2\pi i/3})^{\beta_{\pm}}$ & 
$n^1_{l_1 l_2 L \pm} \equiv l_4+l_7+2(l_5+l_8)$\\
& &~~~~~~~~~~~~~~$= 3- l_1 - 2l_2 - \beta_{\pm}~~~(\mbox{mod}~3)$ \\ \hline
$T^2/Z_4$ & $(-i)^{k}\eta^0_{k \pm}=i^{\alpha_{\pm}}$ & 
$n^0_{l_1 l_2 L \pm} \equiv 2(l_3+l_4)+3(l_5+l_6)$ \\
& &~~~~~~~~~~~~~~$= 4-l_1 -l_2- \alpha_{\pm}~~~(\mbox{mod}~4)$ \\
& $(-1)^{k}\eta^1_{k \pm}=(-1)^{\beta_{\pm}}$ & 
$n^1_{l_1 l_2 L \pm} \equiv l_3+l_5 + l_7 = 2- l_1 - \beta_{\pm}~~~(\mbox{mod}~2)$ \\ \hline
$T^2/Z_6$ & $(e^{-\pi i/3})^{k}\eta^0_{k \pm}=(e^{\pi i/3})^{\alpha_{\pm}}$ & 
$n^0_{l_1 l_2 L \pm} \equiv 2(l_3+l_4)+3(l_5+l_6)$\\
& &~~~~~~~~~~~~~~~~~$+ 4(l_7+l_8) + 5(l_9+l_{10})$\\
& &~~~~~~~~~~~~~~$=6-l_1 -l_2- \alpha_{\pm}~~~(\mbox{mod}~6)$ \\
& $(-1)^{k}\eta^1_{k \pm}=(-1)^{\beta_{\pm}}$ & 
$n^1_{l_1 l_2 L \pm} \equiv l_3+l_5 + l_7 + l_9 + l_{11}$\\
& &~~~~~~~~~~~~~~$= 2- l_1 - \beta_{\pm}~~~(\mbox{mod}~2)$ \\ \hline
\end{tabular}
\end{center}
\end{table}
The relations will be confirmed in the next section.
In the same way, $\sum_{\{l_3, \cdots, l_{n-1}\}_{n^a_{l_1 l_2 R \pm}}}$ means that
the summations over $l_j=0, \cdots, k-l_1-\cdots - l_{j-1}$ $(j=3, \cdots, n-1)$
are carried out under the condition that $l_j$ should satisfy
specific relations $n^a_{l_1 l_2 R \pm}=n^a_{l_1 l_2 L \pm} \mp 1$ (mod $M$) for $\Psi_{\pm}$. 
The formulae (\ref{nd-ZM}) -- (\ref{nnu-ZM}) will be also rewritten in more concrete form
for each $T^2/Z_N$,
by the use of projection operators, in the next section.

\subsection{Generic features of flavor numbers}

We list generic features of flavor numbers.\\
(i) {\it Each flavor number from $[N,k]$ with intrinsic $Z_M$ elements $\eta^a_{k \pm}$
is equal to that from $[N,N-k]$ with appropriate ones $\eta^a_{N-k \pm}$.}

Let us explain this feature using the $SU(5)$ multiplets.
{}From (\ref{Nk(p1=5)}) and the decomposition of $[N, N-k]$ such that
\begin{eqnarray}
[N, N-k] = \sum_{l_1 =0}^{k} \sum_{l_2 = 0}^{k-l_1} \cdots \sum_{l_{n-1} = 0}^{k-l_1-\cdots -l_{n-2}}  
\left({}_{5}C_{5-l_1}, {}_{p_2}C_{p_2-l_2}, \cdots, {}_{p_n}C_{p_n-l_n}\right)~,
\label{NN-k(p1=5)}
\end{eqnarray}
there is a one-to-one correspondence
between $\left({}_{5}C_{5-l_1}, {}_{p_2}C_{p_2-l_2}, \cdots, {}_{p_n}C_{p_n-l_n}\right)$ in $[N, N-k]$
and $\left({}_{5}C_{l_1}, {}_{p_2}C_{l_2}, \cdots, {}_{p_n}C_{l_n}\right)$ in $[N, k]$.
The right-handed Weyl fermion whose representation is 
$\left({}_{5}C_{5-l_1}, {}_{p_2}C_{p_2-l_2}, \cdots, {}_{p_n}C_{p_n-l_n}\right)$ is 
regarded as the left-handed one whose representation is the conjugate representation
$\left({}_{5}C_{l_1}, {}_{p_2}C_{l_2}, \cdots, {}_{p_n}C_{l_n}\right)$,
and hence we obtain the same numbers for (\ref{nbar5-ZM}) -- (\ref{n1-ZM})
with a suitable assignment of intrinsic $Z_M$ elements for $[N, N-k]$.

Here, we give an example for $T^2/Z_2$.
Each flavor number obtained from $[N, k]$
with $(-1)^{k} \eta_{k \pm}^{0} = (-1)^{\alpha_{\pm}}$,
$(-1)^{k} \eta_{k \pm}^{1} = (-1)^{\beta_{\pm}}$ and $(-1)^{k} \eta_{k \pm}^{2} = (-1)^{\gamma_{\pm}}$
agrees with that from $[N, N-k]$ with 
$(-1)^{N-k} \eta_{N-k \pm}^{0} = (-1)^{\alpha'_{\pm}}$,
$(-1)^{N-k} \eta_{N-k \pm}^{1} = (-1)^{\beta'_{\pm}}$ 
and $(-1)^{N-k} \eta_{N-k \pm}^{2} = (-1)^{\gamma'_{\pm}}$,
where $\alpha'_{\pm}$, $\beta'_{\pm}$ and $\gamma'_{\pm}$ satisfy the relations
$\alpha'_{\pm} = \alpha_{\pm} + p_2 + p_3 + p_4 (\mbox{mod} 2)$,
$\beta'_{\pm} = \beta_{\pm} + p_2 + p_5 + p_6 (\mbox{mod} 2)$
and $\gamma'_{\pm} = \gamma_{\pm} + p_3 + p_5 + p_7 (\mbox{mod} 2)$, respectively.\\
(ii) {\it Each flavor number from $[N,k]$ with intrinsic $Z_2$ elements
$(-1)^k \eta^a_{k \pm} = (-1)^{\delta^a_{\pm}}$
is equal to that from $[N,k]$ with the exchanged ones $(\delta^a_+ \leftrightarrow \delta^a_-)$, 
i.e., $(-1)^k \eta^a_{k \pm} = (-1)^{\delta^a_{\mp}}$.}

This feature is understood from the fact that 
specific relations on $l_j$ for $\Psi_{+}$ change into 
those of $\Psi_{-}$ and vice versa, 
under the exchange of $Z_2$ parity of $\Psi_{+}$ and that of $\Psi_{-}$.

Here, we give an example for $T^2/Z_2$.
Under the exchange of $\alpha_+$ and $\alpha_-$, 
$n^0_{l_1 L +}$ and $n^0_{l_1 R +}$ change into $n^0_{l_1 L -}$ and $n^0_{l_1 R -}$
$(\mbox{mod} 2)$, respectively. 
Each flavor number remains the same, because the summation is taken for $\Psi_+$ and $\Psi_-$.\\
(iii) {\it Each flavor number from $[N,k]$ is invariant under several types of exchange 
among $p_j$ and intrinsic $Z_M$ elements.}

From specific relations in Table \ref{T2}, 
we find that the same number for each $SU(5)$ multiplet is obtained under the exchange,
\begin{eqnarray}
&~& (p_3, p_4, \alpha_{\pm}) \Longleftrightarrow (p_5, p_6, \beta_{\pm})~,~~
(p_2, p_6, \beta_{\pm}) \Longleftrightarrow (p_3, p_7, \gamma_{\pm})~,
\nonumber \\
&~& (p_2, p_4, \alpha_{\pm}) \Longleftrightarrow (p_5, p_7, \gamma_{\pm})
~~~~~~~~~~~~~~~~~~~~~~~~~~~~~~~~~~~~~~~~~~~~~~~~~~~~~~~~~~~~~~~~~~ \mbox{for}~~T^2/Z_2~,
\label{ex-Z2}\\
&~& (p_2, p_3, p_6, \alpha_{\pm}) \Longleftrightarrow (p_4, p_7, p_8, \beta_{\pm})
~~~~~~~~~~~~~~~~~~~~~~~~~~~~~~~~~~~~~~~~~~~~~~~~~~~~~ \mbox{for}~~T^2/Z_3~,
\label{ex-Z3}
\end{eqnarray}
where the exchange is done independently.

In the same way, from specific relations in Table \ref{T3}, 
we find that the same number for each SM multiplet is obtained under the exchange,
\begin{eqnarray}
&~& (p_3, p_4, \alpha_{\pm}) \Longleftrightarrow (p_5, p_6, \beta_{\pm})~,~~
~~~~~~~~~~~~~~~~~~~~~~~~~ \mbox{for}~~T^2/Z_2~.
\label{ex-Z2-SM}
\end{eqnarray}

Under the above exchanges, although the unbroken gauge symmetry remains,
the numbers of zero modes for extra-dimensional components of gauge bosons 
are, in general, different 
and hence a model is transformed into a different one.\\
(iv) {\it Each flavor number obtained from $[N,k]$
is invariant in the introduction of Wilson line phases.}

Let us give some examples.

On $T^2/Z_2$, the numbers $n_{\bar{5}}$ and $n_{10}$ obtained from the 
breaking pattern $SU(N) \to  SU(5) \times SU(p_2) \times \cdots \times SU(p_8) \times U(1)^{7-m}$
are same as those from
$SU(N) \to  SU(5) \times SU(p'_2) \times \cdots \times SU(p'_8) \times U(1)^{7-m}$,
if the following relations are satisfied,
\begin{eqnarray}
p'_2 - p_2 = p'_7 - p_7 = p_3 - p'_3 = p_6 - p'_6~,~~p'_4 = p_4~,~~ p'_5 = p_5~,~~ p'_8 = p_8~,
\label{equ-r1}
\end{eqnarray}
or 
\begin{eqnarray}
p'_2 - p_2 = p'_7 - p_7 = p_4 - p'_4 = p_5 - p'_5~,~~p'_3 = p_3~,~~ p'_6 = p_6~,~~ p'_8 = p_8~,
\label{equ-r2}
\end{eqnarray}
or
\begin{eqnarray}
p'_3 - p_3 = p'_6 - p_6 = p_4 - p'_4 = p_5 - p'_5~,~~ p'_2 = p_2~,~~ p'_7 = p_7~,~~p'_8 = p_8~.
\label{equ-r3}
\end{eqnarray}

The above BCs are connected by a singular gauge transformation,
and they are regarded as equivalent in the presence of Wilson line phases.
This equivalence originates from the Hosotani mechanism~\cite{Hosotani1,Hosotani2,HHH&K,HH&K},
and is shown by the following relations among the diagonal representatives 
for $2 \times 2$ submatrices of $(P_0, P_1, P_2)$~\cite{K&M2},
\begin{eqnarray}
(\tau_3, \tau_3, \tau_3) \sim (\tau_3, \tau_3, -\tau_3) \sim (\tau_3, -\tau_3, \tau_3)
\sim (\tau_3, -\tau_3, -\tau_3)~,
\label{equ-Z2}
\end{eqnarray}
where $\tau_3$ is the third component of Pauli matrices.

In our present case, we assume that the BC is chosen as a physical one, i.e., the system with
the physical vacuum is realized with the vanishing Wilson line phases after a suitable gauge transformation
is performed.
Hence, it is understood that each net flavor number obtained from $[N,k]$ does not change   
even though the vacuum changes different ones in the presence of Wilson line phases.

In the same way, the numbers $n_{\bar{d}}$, $n_l$, $n_{\bar{u}}$, $n_{\bar{e}}$ and $n_{q}$ 
obtained from the breaking pattern 
$SU(N) \to SU(3) \times SU(2) \times SU(p_3) \times \cdots \times SU(p_8) \times U(1)^{7-m}$
are same as those from
$SU(N) \to SU(3) \times SU(2) \times SU(p'_3) \times \cdots \times SU(p'_8) \times U(1)^{7-m}$,
if the following relations are satisfied,
\begin{eqnarray}
p'_3 - p_3 = p'_6 - p_6 = p_4 - p'_4 = p_5 - p'_5~,~~ p'_7 = p_7~,~~
p'_8 = p_8~.
\label{equ-r1-SM}
\end{eqnarray}

On $T^2/Z_3$, the numbers $n_{\bar{5}}$ and $n_{10}$ obtained from the 
breaking pattern $SU(N) \to  SU(5) \times SU(p_2) \times \cdots \times SU(p_9) \times U(1)^{8-m}$
are same as those from
$SU(N) \to  SU(5) \times SU(p'_2) \times \cdots \times SU(p'_9) \times U(1)^{8-m}$,
if the following relations are satisfied,
\begin{eqnarray}
p'_2 - p_2 = p'_6 - p_6 = p'_7 - p_7 = p_3 - p'_3 = p_4 - p'_4 = p_8 - p'_8~,~~p'_5 = p_5~,~~p'_9 = p_9~.
\label{equ-r1-Z3}
\end{eqnarray}

The above BCs are also connected by a singular gauge transformation,
and they are regarded as equivalent in the presence of Wilson line phases.
The equivalence is shown using the following relations among the diagonal representatives 
for $3 \times 3$ submatrices of $(\Theta_0, \Theta_1)$ on $T^2/Z_3$~\cite{K&M2},
\begin{eqnarray}
(X, X) \sim (X, \overline{\omega} X) \sim (X, \omega X)~,
\label{equ-Z3}
\end{eqnarray}
where $\omega = e^{2\pi i/3}$, $\overline{\omega} = e^{4\pi i/3}$, 
and $X = \mbox{diag}(1, \omega, \overline{\omega})$.

For these cases, it is also understood that each net flavor number does not change   
even though the vacuum changes different ones in the presence of Wilson line phases.

Although this feature holds for models on $T^2/Z_4$ and $T^2/Z_6$,
there are no examples in our setting,
because of the absence of Wilson line phases changing BCs
but keeping $SU(5)$ or the SM gauge group for $T^2/Z_4$ 
and because of the absence of equivalence relations between diagonal representatives for $T^2/Z_6$~\cite{K&M2}.

\section{Orbifold family unification on $M^4\times T^2/Z_N$}

We investigate the family unification in $SU(N)$ GUTs
for each $T^2/Z_N$ $(N=2, 3, 4, 6)$.

\subsection{Total numbers of models with three families}

Let us present total numbers of models with the three families, for reference.
Total numbers of models with the three families of $SU(5)$ multiplets and the SM multiplets,
which originate from a Dirac fermion whose representation is $[N,k]$ $(k \le N/2)$ of $SU(N)$,
are summarized 
up to $SU(12)$ in Table \ref{T12} and up to $SU(13)$ in Table \ref{T13}, respectively.
In the Tables, the hyphen (-) means no models.
We omit the total numbers of models from $[N, N-k]$, because they agree with those from $[N, k]$,
reflecting the feature (i) in the subsection 2.3.

\begin{table}[htbp]
\caption{Total numbers of models with the three families of $SU(5)$ multiplets.}
\label{T12}
\begin{center}
\begin{tabular}{|c|c|c|c|c|}
\hline                              
&$T^2/Z_2$&$T^2/Z_3$&$T^2/Z_4$&$T^2/Z_6$\\
\hline \hline
\multirow{2}{*}{$SU(8)$} 
&\multirow{2}{*}{-}&[8,3]:24&[8,3]:14&[8,3]:28\\
&&[8,4]:12&[8,4]:16&[8,4]:20\\
\hline
\multirow{2}{*}{$SU(9)$} 
&[9,3]:192&[9,3]:182&[9,3]:142&[9,3]:512 \\
& 
&[9,4]:348&[9,4]:32&[9,4]:800 \\
\hline
\multirow{3}{*}{$SU(10)$} 
&\multirow{3}{*}{-} &[10,3]:852&[10,3]:160&[10,3]:2484 \\
&&[10,4]:1308&[10,4]:92&[10,4]:2654 \\
&&[10,5]:48& 
&[10,5]:1532 \\
\hline
\multirow{3}{*}{$SU(11)$} 
&[11,3]:768&[11,3]:1608&[11,3]:456&[11,3]:6530 \\
&[11,4]:768&[11,4]:1716&[11,4]:436&[11,4]:6768 \\
& 
&[11,5]:1794&[11,5]:186&[11,5]:5540 \\
\hline
\multirow{4}{*}{$SU(12)$} 
&[12,3]:1104&[12,3]:2214&[12,3]:748&[12,3]:17084 \\
& 
&[12,4]:1020&[12,4]:676&[12,4]:13692 \\
&& 
&[12,5]:534&[12,5]:10498  \\
&& 
&[12,6]:632&[12,6]:13188  \\
\hline
\end{tabular}
\end{center}
\end{table}


\begin{table}[htbp]
\caption{Total numbers of models with the three families of SM multiplets.}
\label{T13}
\begin{center}
\begin{tabular}{|c|c|c|c|c|}
\hline                          
&$T^2/Z_2$&$T^2/Z_3$&$T^2/Z_4$&$T^2/Z_6$\\
\hline \hline
$SU(8)$&-&-&-&-\\
\hline
\multirow{2}{*}{$SU(9)$} 
&[9,3]:32&\multirow{2}{*}{-}&[9,3]:8&[9,3]:8 \\
& 
&& &[9,4]:32 \\
\hline
\multirow{2}{*}{$SU(10)$} 
&\multirow{2}{*}{-}&\multirow{2}{*}{-}&\multirow{2}{*}{-}&[10,3]:80 \\
&&&&[10,4]:108 \\
\hline
\multirow{3}{*}{$SU(11)$} 
&[11,3]:80&[11,4]:80&[11,3]:20&[11,3]:84 \\
&[11,4]:80& 
&[11,4]:20&[11,4]:144 \\
& 
&& &[11,5]:156 \\
\hline
\multirow{4}{*}{$SU(12)$} 
&[12,3]:120&[12,3]:80&[12,4]:88&[12,3]:392 \\
& 
& &[12,6]:240&[12,4]:120 \\
&&& 
&[12,5]:72 \\
&&&&[12,6]:552 \\ 
\hline
\multirow{4}{*}{$SU(13)$} 
&[13,3]:144&\multirow{4}{*}{-}&[13,4]:40&[13,3]:712 \\
& 
&& &[13,4]:88 \\
&&&&[13,5]:140 \\
&&&&[13,6]:200 \\
\hline
\end{tabular}
\end{center}
\end{table}

\subsection{$T^2/Z_2$}

For the representation matrices given by
\begin{eqnarray}
\hspace{-1.5cm}&~& P_{0} = {\mbox{diag}}([+1]_{p_1},[+1]_{p_2},[+1]_{p_3},[+1]_{p_4},[-1]_{p_5},[-1]_{p_6},
[-1]_{p_7},[-1]_{p_8})~,
\nonumber \\
\hspace{-1.5cm}&~& P_{1} = {\mbox{diag}}([+1]_{p_1},[+1]_{p_2},[-1]_{p_3},[-1]_{p_4},[+1]_{p_5},[+1]_{p_6},
[-1]_{p_7},[-1]_{p_8})~,
\nonumber \\
\hspace{-1.5cm}&~& P_{2} = {\mbox{diag}}([+1]_{p_1},[-1]_{p_2},[+1]_{p_3},[-1]_{p_4},[+1]_{p_5},[-1]_{p_6},
[+1]_{p_7},[-1]_{p_8})~,
\label{Z2-P}
\end{eqnarray}
the following breakdown of $SU(N)$ gauge symmetry occurs
\begin{eqnarray}
SU(N) \to SU(p_1)\times SU(p_2) \times \cdots \times SU(p_8)\times U(1)^{7-n}~,
\label{Z2-GSB}
\end{eqnarray}
where $[\pm 1]_{p_i}$ represents $\pm 1$ for all $p_i$ elements.

After the breakdown of $SU(N)$, $[N, k]_{\pm}$
is decomposed as
\begin{eqnarray}
[N, k]_{\pm} = \sum_{l_1 =0}^{k} \sum_{l_2 = 0}^{k-l_1} \cdots \sum_{l_7 = 0}^{k-l_1-\cdots -l_6}  
\left({}_{p_1}C_{l_1}, {}_{p_2}C_{l_2}, \cdots, {}_{p_8}C_{l_8}\right)_{\pm}~,
\label{Nk-Z2}
\end{eqnarray}
where $l_8=k-l_1- \cdots -l_7$.

Using the definition of the intrinsic $Z_2$ parities $\eta^a_{k \pm}$ $(a=0, 1, 2)$ such that
\begin{eqnarray}
({\bm{N}} \times \dots \times {\bm{N}})_{\tiny {\mbox{A}} \pm} 
\to \eta^a_{k \pm} (P_a {\bm{N}} \times \dots \times P_a {\bm{N}})_{\tiny {\mbox{A}} \pm}~,
\label{etaNka-Z2}
\end{eqnarray}
the $Z_2$ parities of the representation 
$\left({}_{p_1}C_{l_1}, {}_{p_2}C_{l_2}, \cdots, {}_{p_8}C_{l_8}\right)_{\pm}$ 
are given by
\begin{eqnarray}
\hspace{-1cm}&~& \mathcal{P}_{0 \pm} = (-1)^{l_5+l_6+l_7+l_8} \eta^0_{k \pm} 
= (-1)^{l_1+l_2+l_3+l_4} (-1)^k \eta^0_{k \pm}= (-1)^{l_1+l_2+l_3+l_4+\alpha_{\pm}}~, 
\label{Z20}\\
\hspace{-1cm}&~& \mathcal{P}_{1 \pm} = (-1)^{l_3+l_4+l_7+l_8} \eta^1_{k \pm} 
= (-1)^{l_1+l_2+l_5+l_6} (-1)^k \eta^1_{k \pm}= (-1)^{l_1+l_2+l_5+l_6+\beta_{\pm}}~,
\label{Z21}\\
\hspace{-1cm}&~& \mathcal{P}_{2 \pm} = (-1)^{l_2+l_4+l_6+l_8} \eta^2_{k \pm} 
= (-1)^{l_1+l_3+l_5+l_7} (-1)^k \eta^2_{k \pm} = (-1)^{l_1+l_3+l_5+l_7+\gamma_{\pm}}~,
\label{Z22}
\end{eqnarray}
where $\eta^a_{k \pm}$ take a value $+1$ or $-1$ by definition,
and we parameterize them as $(-1)^{k}\eta^0_{k \pm} = (-1)^{\alpha_{\pm}}$, 
$(-1)^{k}\eta^1_{k \pm} = (-1)^{\beta_{\pm}}$ and $(-1)^{k}\eta^2_{k \pm} = (-1)^{\gamma_{\pm}}$.

\subsubsection{Numbers of $SU(5)$ multiplets on $T^2/Z_2$}

After the breakdown $SU(N) \to  SU(5) \times SU(p_2) \times \cdots \times SU(p_8) \times U(1)^{7-m}$, 
$[N, k]_{\pm}$ is decomposed as
\begin{eqnarray}
[N, k]_{\pm} = \sum_{l_1 =0}^{k} \sum_{l_2 = 0}^{k-l_1} \cdots \sum_{l_7 = 0}^{k-l_1-\cdots -l_6}  
\left({}_{5}C_{l_1}, {}_{p_2}C_{l_2}, \cdots, {}_{p_8}C_{l_8}\right)_{\pm}~.
\label{Nk(p1=5)Z2}
\end{eqnarray}

Using the assignment of $Z_2$ parities (\ref{Z20}) -- (\ref{Z22}),
we find that zero modes appear if the following relations are satisfied,
\begin{eqnarray}
&~& n^0_{l_1 L \pm} \equiv l_2+l_3+l_4 = 2-l_1 - \alpha_{\pm}~~~(\mbox{mod}~2)~,~~
\nonumber \\
&~& n^1_{l_1 L \pm} \equiv l_2+l_5+l_6 = 2-l_1 - \beta_{\pm}~~~(\mbox{mod}~2)~,~~
\nonumber \\
&~& n^2_{l_1 L \pm} \equiv l_3+l_5+l_7 = 2 - l_1 - \gamma_{\pm} ~~~(\mbox{mod}~2)~.
\label{sum-Z2}
\end{eqnarray}

Utilizing the survival hypothesis and the equivalence of charge conjugation,
we obtain the formulae (\ref{nbar5-ZM}) -- (\ref{n1-ZM}) with $n=8$.
Because the $Z_2$ projection operator $P_{\pm}$ that picks up $\mathcal{P} = \pm 1$ is defined as
$P_{\pm} \equiv (1 \pm \mathcal{P})/2$,
the $Z_2$ projection operator that picks up zero modes of left-handed ones, i.e.,   
massless modes in fields with $(\mathcal{P}_{0 \pm}, \mathcal{P}_{1 \pm}, \mathcal{P}_{2 \pm})=(1, 1, 1)$,
is given by
\begin{eqnarray}
P^{(1,1,1)} \equiv \frac{1}{8}(1 + \mathcal{P}_{0 \pm})(1 + \mathcal{P}_{1 \pm})(1 + \mathcal{P}_{2 \pm})~,
\label{P+1}
\end{eqnarray}
and the $Z_2$ projection operator that picks up the zero modes of right-handed ones, i.e.,   
massless modes in fields with $(\mathcal{P}_{0 \pm}, \mathcal{P}_{1 \pm}, \mathcal{P}_{2 \pm})=(-1, -1, -1)$,
is given by
\begin{eqnarray}
P^{(-1,-1,-1)} \equiv \frac{1}{8}(1 - \mathcal{P}_{0 \pm})(1 - \mathcal{P}_{1 \pm})(1 - \mathcal{P}_{2 \pm})~.
\label{P-1}
\end{eqnarray}
From (\ref{P+1}) and (\ref{P-1}),
\begin{eqnarray}
&~& P^{(1,1,1)} - P^{(-1,-1,-1)} = \frac{1}{4}\left(\mathcal{P}_{0 \pm} + \mathcal{P}_{1 \pm} + \mathcal{P}_{2 \pm}
+ \mathcal{P}_{0 \pm}\mathcal{P}_{1 \pm}\mathcal{P}_{2 \pm}\right)~,
\label{P+P}\\
&~& P^{(1,1,1)} + P^{(-1,-1,-1)} = \frac{1}{4}\left(1+ \mathcal{P}_{0 \pm}\mathcal{P}_{1 \pm} + \mathcal{P}_{0 \pm}\mathcal{P}_{2 \pm}
+ \mathcal{P}_{1 \pm}\mathcal{P}_{2 \pm}\right)~.
\label{P-P}
\end{eqnarray}
Using (\ref{Z20}), (\ref{Z21}), (\ref{Z22}), (\ref{P+P}) and (\ref{P-P}), 
the formulae (\ref{nbar5-ZM}) -- (\ref{n1-ZM}) are rewritten as
\begin{eqnarray}
\hspace{-1cm}&~& n_{\bar{5}} 
= \sum_{\pm} \sum_{l_1 = 1, 4} \sum_{l_2=0}^{k-l_1} \cdots \sum_{l_7=0}^{k-l_1-\cdots -l_6}
(-1)^{l_1} \left(P^{(1,1,1)} - P^{(-1,-1,-1)}\right)
{}_{p_2}C_{l_2} \cdots {}_{p_{8}}C_{l_{8}}
\nonumber \\
\hspace{-1cm}&~& ~~~~~
= \sum_{\pm} \sum_{l_1 = 1, 4} \sum_{l_2=0}^{k-l_1} \cdots \sum_{l_7=0}^{k-l_1-\cdots -l_6}
\frac{1}{4}\left((-1)^{l_2+l_3+l_4 +\alpha_{\pm}} + (-1)^{l_2+l_5+l_6 + \beta_{\pm}} \right.
\nonumber \\
\hspace{-1cm}&~& ~~~~~~~~~~~~~~~~~~~~~~~~~~~~~ \left. + (-1)^{l_3+l_5+l_7+\gamma_{\pm}} 
+ (-1)^{l_4+l_6+l_7 +\alpha_{\pm}+\beta_{\pm}+\gamma_{\pm}} \right)
{}_{p_2}C_{l_2} \cdots {}_{p_{8}}C_{l_{8}}~, 
\label{nbar5-Z2-P}\\
\hspace{-1cm}&~& n_{10} 
= \sum_{\pm} \sum_{l_1 = 2, 3} \sum_{l_2=0}^{k-l_1} \cdots \sum_{l_7=0}^{k-l_1-\cdots -l_6}
(-1)^{l_1} \left(P^{(1,1,1)} - P^{(-1,-1,-1)}\right)
{}_{p_2}C_{l_2} \cdots {}_{p_{8}}C_{l_{8}}
\nonumber \\
\hspace{-1cm}&~& ~~~~~~
= \sum_{\pm} \sum_{l_1 = 2, 3} \sum_{l_2=0}^{k-l_1} \cdots \sum_{l_7=0}^{k-l_1-\cdots -l_6}
\frac{1}{4}\left((-1)^{l_2+l_3+l_4 +\alpha_{\pm}} + (-1)^{l_2+l_5+l_6 + \beta_{\pm}} \right.
\nonumber \\
\hspace{-1cm}&~& ~~~~~~~~~~~~~~~~~~~~~~~~~~~~~ \left. + (-1)^{l_3+l_5+l_7+\gamma_{\pm}} 
+ (-1)^{l_4+l_6+l_7 +\alpha_{\pm}+\beta_{\pm}+\gamma_{\pm}} \right)
{}_{p_2}C_{l_2} \cdots {}_{p_{8}}C_{l_{8}}~, 
\label{n10-Z2-P}\\
\hspace{-1cm}&~& n_{1} 
= \sum_{\pm} \sum_{l_1 = 0, 5} \sum_{l_2=0}^{k-l_1} \cdots \sum_{l_7=0}^{k-l_1-\cdots -l_6}
\left(P^{(1,1,1)} + P^{(-1,-1,-1)}\right)
{}_{p_2}C_{l_2} \cdots {}_{p_{8}}C_{l_{8}}
\nonumber \\
\hspace{-1cm}&~& ~~~~~
= \sum_{\pm} \sum_{l_1 = 0, 5} \sum_{l_2=0}^{k-l_1} \cdots \sum_{l_7=0}^{k-l_1-\cdots -l_6}
\frac{1}{4}\left(1 + (-1)^{l_3+l_4+l_5+l_6 +\alpha_{\pm}+\beta_{\pm}} \right.
\nonumber \\
\hspace{-1cm}&~& ~~~~~~~~~~~~~~~~~~~ \left. + (-1)^{l_2 + l_4+l_5+l_7+\alpha_{\pm}+\gamma_{\pm}} 
+ (-1)^{l_2+l_3+l_6+l_7 +\beta_{\pm}+\gamma_{\pm}} \right)
{}_{p_2}C_{l_2} \cdots {}_{p_{8}}C_{l_{8}}~.
\label{n1-Z2-P}
\end{eqnarray}

Here, we give some examples for representations and BCs to derive $n_{\bar{5}} = n_{10} = 3$, in Table \ref{T4}. 
\begin{table}[htbp]
\caption{Examples for the three families of $SU(5)$ from $T^2/Z_2$.}
\label{T4}
\begin{center}
\begin{tabular}{|c|c|c|c|}  
\hline
$[N,k]$&$(p_1,p_2,p_3,p_4,p_5,p_6,p_7,p_8)$&$(\alpha_+,\beta_+,\gamma_+)$&$(\alpha_-,\beta_-,\gamma_-)$\\
\hline \hline
[9,3]&(5,0,0,0,3,0,0,1)&(0,1,1)&(0,0,1)\\
\hline
[11,3]&(5,0,1,0,4,0,1,0)&(0,0,1)&(1,1,0)\\
\hline
[11,4]&(5,0,3,1,0,1,1,0)&(0,0,0)&(0,0,1)\\
\hline
[12,3]&(5,2,0,0,2,0,1,2)&(1,0,1)&(0,0,0)\\
\hline
\end{tabular}
\end{center}
\end{table}

\subsubsection{Numbers of the SM multiplets on $T^2/Z_2$}

After the breakdown $SU(N) \to  SU(3) \times SU(2) \times SU(p_2) \times \cdots \times SU(p_8) \times U(1)^{7-m}$, 
$[N, k]_{\pm}$ is decomposed as
\begin{eqnarray}
[N, k]_{\pm} 
= \sum_{l_1 =0}^{k} \sum_{l_2 = 0}^{k-l_1} \sum_{l_3 = 0}^{k-l_1-l_2} \cdots \sum_{l_7 = 0}^{k-l_1-\cdots -l_6}  
\left({}_{3}C_{l_1}, {}_{2}C_{l_2}, {}_{p_3}C_{l_3}, \cdots, {}_{p_8}C_{l_8}\right)_{\pm}~.
\label{Nk(p1=3)Z2}
\end{eqnarray}

Using the assignment of $Z_2$ parities (\ref{Z20}) -- (\ref{Z22}),
we find that zero modes appear if the following relations are satisfied,
\begin{eqnarray}
&~& n^0_{l_1 l_2 L \pm} \equiv l_3+l_4 = 2-l_1 - l_2 - \alpha_{\pm}~~~(\mbox{mod}~2)~,~~
\nonumber \\
&~& n^1_{l_1 l_2 L \pm} \equiv l_5+l_6 = 2-l_1 - l_2 - \beta_{\pm}~~~(\mbox{mod}~2)~,~~
\nonumber \\
&~& n^2_{l_1 l_2 L \pm} \equiv l_3+l_5+l_7 = 2 - l_1 - \gamma_{\pm} ~~~(\mbox{mod}~2)~,
\label{sum-Z2-SM}
\end{eqnarray}
for $(-1)^{k}\eta^0_{k \pm} = (-1)^{\alpha_{\pm}}$, 
$(-1)^{k}\eta^1_{k \pm} = (-1)^{\beta_{\pm}}$ 
and $(-1)^{k}\eta^2_{k \pm} = (-1)^{\gamma_{\pm}}$.

Then, we obtain the formulae (\ref{nd-ZM}) -- (\ref{nnu-ZM}) with $n=8$.
Using (\ref{Z20}), (\ref{Z21}), (\ref{Z22}), (\ref{P+P}) and (\ref{P-P}), 
the formulae for $(d_R)^c$ and $(\nu_R)^c$ are rewritten as
\begin{eqnarray}
\hspace{-1cm}&~& n_{\bar{d}} = \sum_{\pm} \sum_{(l_1, l_2) = (2,2),(1,0)}
\sum_{l_2=0}^{k-l_1} \cdots \sum_{l_7=0}^{k-l_1-\cdots -l_6}
\frac{1}{4}\left((-1)^{l_3+l_4 +\alpha_{\pm}} + (-1)^{l_5+l_6 + \beta_{\pm}} \right.
\nonumber \\
\hspace{-1cm}&~& ~~~~~~~~~~~~~~~~~~~~~~ \left. + (-1)^{l_2 + l_3+l_5+l_7+\gamma_{\pm}} 
+ (-1)^{l_2+l_4+l_6+l_7 +\alpha_{\pm}+\beta_{\pm}+\gamma_{\pm}} \right)
{}_{p_2}C_{l_2} \cdots {}_{p_{8}}C_{l_{8}}~, 
\label{nd-Z2-P}\\
\hspace{-1cm}&~& n_{\bar{\nu}} = \sum_{\pm} \sum_{(l_1, l_2) = (0,0),(3,2)} 
 \sum_{l_2=0}^{k-l_1} \cdots \sum_{l_7=0}^{k-l_1-\cdots -l_6}
\frac{1}{4}\left(1 + (-1)^{l_3+l_4+l_5+l_6 +\alpha_{\pm}+\beta_{\pm}} \right.
\nonumber \\
\hspace{-1cm}&~& ~~~~~~~~~~~~~~~~~~~ \left. + (-1)^{l_2 + l_4+l_5+l_7+\alpha_{\pm}+\gamma_{\pm}} 
+ (-1)^{l_2+l_3+l_6+l_7 +\beta_{\pm}+\gamma_{\pm}} \right)
{}_{p_2}C_{l_2} \cdots {}_{p_{8}}C_{l_{8}}~.
\label{nnu-Z2-P}
\end{eqnarray}
The formulae for $l_L$, $(u_R)^c$, $(e_R)^c$ and $q_L$ are
obtained by replacing the summation of $(l_1, l_2)$ for $n_{\bar{d}}$
with $\{(3,1),(0,1)\}$, $\{(2,0),(1,2)\}$, $\{(0,2),(3,0)\}$
and $\{(1,1),(2,1)\}$. 

Here, we give a list of all BCs to derive three families of SM fermions
from $[9,3]$, in Table \ref{T5}. 
We find that the features (ii) and (iii), presented in subsection 2.3, hold on.
\begin{table}[htbp]
\caption{The three families of SM multiplets from $[9,3]$ on $T^2/Z_2$.}
\label{T5}
\begin{center}
\begin{tabular}{|c|c|c|c|}  
\hline
$[N,k]$&$(p_1,p_2,p_3,p_4,p_5,p_6,p_7,p_8)$&$(\alpha_+,\beta_+,\gamma_+)$&$(\alpha_-,\beta_-,\gamma_-)$\\
\hline\hline
\multirow{32}{*}{[9,3]}
&(3,2,0,0,0,3,0,1)&(0,1,1)&(0,1,0)\\
&(3,2,0,0,0,3,0,1)&(0,1,0)&(0,1,1)\\
&(3,2,0,0,0,3,1,0)&(0,1,1)&(0,1,0)\\
&(3,2,0,0,0,3,1,0)&(0,1,0)&(0,1,1)\\
&(3,2,0,0,3,0,0,1)&(0,1,1)&(0,1,0)\\
&(3,2,0,0,3,0,0,1)&(0,1,0)&(0,1,1)\\
&(3,2,0,0,3,0,1,0)&(0,1,1)&(0,1,0)\\
&(3,2,0,0,3,0,1,0)&(0,1,0)&(0,1,1)\\
&(3,2,0,3,0,0,0,1)&(1,0,1)&(1,0,0)\\
&(3,2,0,3,0,0,0,1)&(1,0,0)&(1,0,1)\\
&(3,2,0,3,0,0,1,0)&(1,0,1)&(1,0,0)\\
&(3,2,0,3,0,0,1,0)&(1,0,0)&(1,0,1)\\
&(3,2,3,0,0,0,0,1)&(1,0,1)&(1,0,0)\\
&(3,2,3,0,0,0,0,1)&(1,0,0)&(1,0,1)\\
&(3,2,3,0,0,0,1,0)&(1,0,1)&(1,0,0)\\
&(3,2,3,0,0,0,1,0)&(1,0,0)&(1,0,1)\\
&(3,2,0,0,1,2,0,1)&(0,1,1)&(0,1,0)\\
&(3,2,0,0,1,2,0,1)&(0,1,0)&(0,1,1)\\
&(3,2,0,0,1,2,1,0)&(0,1,1)&(0,1,0)\\
&(3,2,0,0,1,2,1,0)&(0,1,0)&(0,1,1)\\
&(3,2,0,0,2,1,0,1)&(0,1,1)&(0,1,0)\\
&(3,2,0,0,2,1,0,1)&(0,1,0)&(0,1,1)\\
&(3,2,0,0,2,1,1,0)&(0,1,1)&(0,1,0)\\
&(3,2,0,0,2,1,1,0)&(0,1,0)&(0,1,1)\\
&(3,2,1,2,0,0,0,1)&(1,0,1)&(1,0,0)\\
&(3,2,1,2,0,0,0,1)&(1,0,0)&(1,0,1)\\
&(3,2,1,2,0,0,1,0)&(1,0,1)&(1,0,0)\\
&(3,2,1,2,0,0,1,0)&(1,0,0)&(1,0,1)\\
&(3,2,2,1,0,0,0,1)&(1,0,1)&(1,0,0)\\
&(3,2,2,1,0,0,0,1)&(1,0,0)&(1,0,1)\\
&(3,2,2,1,0,0,1,0)&(1,0,1)&(1,0,0)\\
&(3,2,2,1,0,0,1,0)&(1,0,0)&(1,0,1)\\
\hline
\end{tabular}
\end{center}
\end{table}

\subsection{$T^2/Z_3$}

For the representation matrices given by
\begin{eqnarray}
\hspace{-1.5cm}&~& \Theta_{0} = {\mbox{diag}}([1]_{p_1},[1]_{p_2},[1]_{p_3},[\omega]_{p_4},[\omega]_{p_5},[\omega]_{p_6},
[\overline{\omega}]_{p_7},[\overline{\omega}]_{p_8},[\overline{\omega}]_{p_9})~,
\nonumber \\
\hspace{-1.5cm}&~& \Theta_{1} = {\mbox{diag}}([1]_{p_1},[\omega]_{p_2},[\overline{\omega}]_{p_3},
[1]_{p_4},[\omega]_{p_5},[\overline{\omega}]_{p_6},
[1]_{p_7},[{\omega}]_{p_8},[\overline{\omega}]_{p_9})~,
\label{Z3-Theta}
\end{eqnarray}
the following breakdown of $SU(N)$ gauge symmetry occurs
\begin{eqnarray}
SU(N) \to SU(p_1)\times SU(p_2) \times \cdots \times SU(p_9)\times U(1)^{7-n}~,
\label{Z3-GSB}
\end{eqnarray}
where $[1]_{p_i}$, $[\omega]_{p_i}$ and $[\overline{\omega}]_{p_i}$ represent
$1$, $\omega (\equiv e^{2\pi i/3})$ and $\overline{\omega}(\equiv e^{4\pi i/3})$
for all $p_i$ elements.

After the breakdown of $SU(N)$, $[N, k]_{\pm}$
is decomposed as
\begin{eqnarray}
[N, k]_{\pm} = \sum_{l_1 =0}^{k} \sum_{l_2 = 0}^{k-l_1} \cdots \sum_{l_8 = 0}^{k-l_1-\cdots -l_7}  
\left({}_{p_1}C_{l_1}, {}_{p_2}C_{l_2}, \cdots, {}_{p_9}C_{l_9}\right)_{\pm}~,
\label{Nk-Z3}
\end{eqnarray}
where $l_9=k-l_1- \cdots -l_8$.
The $\left({}_{p_1}C_{l_1}, {}_{p_2}C_{l_2}, \cdots, {}_{p_9}C_{l_9}\right)_{\pm}$ 
has the $Z_3$ elements 
\begin{eqnarray}
&~& \mathcal{P}_{0 \pm} = \omega^{l_4+l_5+l_6} \overline{\omega}^{l_7+l_8+l_9} \eta^0_{k \pm}
= \omega^{l_1+l_2+l_3+ 2(l_4+l_5+l_6)} \overline{\omega}^{k} \eta^0_{k \pm}
\nonumber \\
&~& ~~~~~~~~~
= \omega^{l_1+l_2+l_3+ 2(l_4+l_5+l_6) + \alpha_{\pm}} ~, 
\label{Z30}\\
&~& \mathcal{P}_{1 \pm} = \omega^{l_2+l_5+l_8} \overline{\omega}^{l_3+l_6+l_9} \eta^1_{k \pm}
= \omega^{l_1+l_4+l_7+ 2(l_2+l_5+l_8)} \overline{\omega}^{k} \eta^1_{k \pm}
\nonumber \\
&~& ~~~~~~~~~
= \omega^{l_1+l_4+l_7+ 2(l_2+l_5+l_8) + \beta_{\pm}}~,
\label{Z31}
\end{eqnarray}
where $\eta^a_{k \pm}$ take a value $1$, $\omega$ or $\overline{\omega}$,
and we parameterize them as $\overline{\omega}^k \eta^0_{k \pm}= \omega^{\alpha_{\pm}}$ 
and $\overline{\omega}^k \eta^1_{k \pm}= \omega^{\beta_{\pm}}$.

\subsubsection{Numbers of $SU(5)$ multiplets on $T^2/Z_3$}

After the breakdown of $SU(N) \to SU(5) \times SU(p_2) \times \cdots \times SU(p_9) \times U(1)^{8-m}$, 
$[N, k]_{\pm}$ is decomposed as
\begin{eqnarray}
[N, k]_{\pm} = \sum_{l_1 =0}^{k} \sum_{l_2 = 0}^{k-l_1} \cdots \sum_{l_8 = 0}^{k-l_1-\cdots -l_7}  
\left({}_{5}C_{l_1}, {}_{p_2}C_{l_2}, \cdots, {}_{p_9}C_{l_9}\right)_{\pm}~.
\label{Nk(p1=5)Z3}
\end{eqnarray}

Using the assignment of $Z_3$ elements (\ref{Z30}) and (\ref{Z31}),
we find that zero modes appear if the following relations are satisfied,
\begin{eqnarray}
&~& n^0_{l_1 L \pm} \equiv l_2+l_3+2(l_4+l_5+l_6) = 3-l_1 - \alpha_{\pm}~~~(\mbox{mod}~3)~,~~
\nonumber \\
&~& n^1_{l_1 L \pm} \equiv l_4+l_7+2(l_2+l_5+l_8) = 3- l_1 - \beta_{\pm}~~~(\mbox{mod}~3)~.
\label{sum-Z3L}
\end{eqnarray}
The relation $n^a_{l_1 R \pm}=n^a_{l_1 L \pm} \mp 1$ (mod $3$) holds from (\ref{rhoPsiR}).

Then, we obtain the formulae (\ref{nbar5-ZM}) -- (\ref{n1-ZM}) with $n=9$,
and they are rewritten as
\begin{eqnarray}
\hspace{-1cm}&~& n_{\bar{5}} = \sum_{l_1 = 1, 4} \sum_{l_2=0}^{k-l_1} \cdots \sum_{l_8=0}^{k-l_1-\cdots -l_7}
(-1)^{l_1} 
\left(P^{(1,1)}_{+} - P^{(\omega, \omega)}_{+} 
+ P^{(1,1)}_{-} - P^{(\overline{\omega}, \overline{\omega})}_{-}\right)
{}_{p_2}C_{l_2} \cdots {}_{p_{9}}C_{l_{9}}~, 
\label{nbar5-Z3-P}\\
\hspace{-1cm}&~& n_{10} = \sum_{l_1 = 2, 3} \sum_{l_2=0}^{k-l_1} \cdots \sum_{l_8=0}^{k-l_1-\cdots -l_7}
(-1)^{l_1} 
\left(P^{(1,1)}_{+} - P^{(\omega, \omega)}_{+} 
+ P^{(1,1)}_{-} - P^{(\overline{\omega}, \overline{\omega})}_{-}\right)
{}_{p_2}C_{l_2} \cdots {}_{p_{9}}C_{l_{9}}~, 
\label{n10-Z3-P}\\
\hspace{-1cm}&~& n_{1} = \sum_{l_1 = 0, 5} \sum_{l_2=0}^{k-l_1} \cdots \sum_{l_8=0}^{k-l_1-\cdots -l_7}
\left(P^{(1,1)}_{+} + P^{(\omega, \omega)}_{+} 
+ P^{(1,1)}_{-} + P^{(\overline{\omega}, \overline{\omega})}_{-}\right) {}_{p_2}C_{l_2} \cdots {}_{p_{9}}C_{l_{9}}~,
\label{n1-Z3-P}
\end{eqnarray}
where $P^{(\rho, \rho)}_{\pm}$ are projection operators that pick up the part relating
$(\mathcal{P}_{0 \pm}, \mathcal{P}_{1 \pm})=(\rho, \rho)$ and are written by
\begin{eqnarray}
P^{(\rho, \rho)}_{\pm} = \frac{1}{9}
\left(1+ \overline{\rho} \mathcal{P}_{0 \pm} + \overline{\rho}^2 \mathcal{P}_{0 \pm}^2\right)
\left(1+ \overline{\rho} \mathcal{P}_{1 \pm} + \overline{\rho}^2 \mathcal{P}_{1 \pm}^2\right)~.
\label{P-Z3}
\end{eqnarray}

Here, we give some examples for representations and BCs to derive $n_{\bar{5}} = n_{10} = 3$, in Table \ref{T6}. 
\begin{table}[htbp]
\caption{Examples for the three families of $SU(5)$ from $T^2/Z_3$.}
\label{T6}
\begin{center}
\begin{tabular}{|c|c|c|c|}  
\hline
$[N,k]$&$(p_1,p_2,p_3,p_4,p_5,p_6,p_7,p_8,p_9)$&$(\alpha_+,\beta_+)$&$(\alpha_-,\beta_-)$\\
\hline \hline
[8,3]&(5,0,0,0,3,0,0,0,0)&(2,0)&(2,2) \\
\hline
[8,4]&(5,1,1,0,1,0,0,0,0)&(0,0)&(2,2) \\
\hline
[9,3]&(5,0,0,2,0,1,0,0,1)&(2,0)&(2,1) \\
\hline
[9,4]&(5,0,2,0,0,0,0,2,0)&(2,2)&(0,2) \\
\hline
[10,3]&(5,0,0,0,3,2,0,0,0)&(2,0)&(2,2) \\
\hline
[10,4]&(5,0,0,1,0,1,1,1,1)&(2,2)&(2,2) \\
\hline
[10,5]&(5,1,0,0,1,0,2,0,1)&(0,0)&(0,0) \\
\hline
[11,3]&(5,1,0,0,1,4,0,0,0)&(0,0)&(2,1) \\
\hline
[11,4]&(5,2,2,0,0,1,0,1,0)&(1,2)&(2,1) \\
\hline
[11,5]&(5,1,1,1,1,0,0,0,2)&(0,1)&(1,1) \\
\hline
[12,3]&(5,0,0,3,3,0,0,0,1)&(2,0)&(0,2) \\
\hline
[12,4]&(5,0,3,1,0,1,0,2,0)&(1,2)&(0,1) \\
\hline
\end{tabular}
\end{center}
\end{table}

\subsubsection{Numbers of the SM multiplets on $T^2/Z_3$}

After the breakdown $SU(N) \to  SU(3) \times SU(2) \times SU(p_2) \times \cdots \times SU(p_9) \times U(1)^{8-m}$, 
$[N, k]_{\pm}$ is decomposed as
\begin{eqnarray}
[N, k]_{\pm} = \sum_{l_1 =0}^{k} \sum_{l_2 = 0}^{k-l_1} \sum_{l_3 = 0}^{k-l_1-l_2} \cdots \sum_{l_8 = 0}^{k-l_1-\cdots -l_7}  
\left({}_{3}C_{l_1}, {}_{2}C_{l_2}, {}_{p_3}C_{l_3}, \cdots, {}_{p_9}C_{l_9}\right)_{\pm}~.
\label{Nk(p1=3)Z3}
\end{eqnarray}

Using the assignment of $Z_3$ elements (\ref{Z30}) and (\ref{Z31}),
we find that zero modes appear if the following relations are satisfied,
\begin{eqnarray}
&~& n^0_{l_1 l_2 L \pm} \equiv l_3+2(l_4+l_5+l_6) = 3-l_1 -l_2- \alpha_{\pm}~~~(\mbox{mod}~3)~,~~
\nonumber \\
&~& n^1_{l_1 l_2 L \pm} \equiv l_4+l_7+2(l_5+l_8) = 3- l_1 -2l_2- \beta_{\pm}~~~(\mbox{mod}~3)~.
\label{sum-Z3L-SM}
\end{eqnarray}
The relation $n^a_{l_1 l_2 R \pm}=n^a_{l_1 l_2 L \pm}\mp 1$ (mod $3$) 
holds from (\ref{rhoPsiR}).

Then, we obtain the formulae (\ref{nd-ZM}) -- (\ref{nnu-ZM}) with $n=9$.
Using the projection operators (\ref{P-Z3}), 
the formulae for $(d_R)^c$ and $(\nu_R)^c$ are rewritten as
\begin{eqnarray}
\hspace{-1.3cm}&~& n_{\bar{d}} = \!\!\!\!\! \sum_{(l_1, l_2) = (2,2),(1,0)}
\sum_{l_2=0}^{k-l_1} \cdots \sum_{l_8=0}^{k-l_1-\cdots -l_7}
\!\!\!\!\!\!\!\!\!(-1)^{l_1+l_2} 
 \left(P^{(1,1)}_{+} 
- P^{(\omega, \omega)}_{+} + P^{(1,1)}_{-} - P^{(\overline{\omega}, \overline{\omega})}_{-}\right)
{}_{p_2}C_{l_2} \cdots {}_{p_{9}}C_{l_{9}}~, 
\label{nd-Z3-P}\\
\hspace{-1.3cm}&~& n_{\bar{\nu}} = \!\!\!\!\! \sum_{(l_1, l_2) = (0,0),(3,2)} 
 \sum_{l_2=0}^{k-l_1} \cdots \sum_{l_8=0}^{k-l_1-\cdots -l_7}
\left(P^{(1,1)}_{+} 
+ P^{(\omega, \omega)}_{+} + P^{(1,1)}_{-} + P^{(\overline{\omega}, \overline{\omega})}_{-}\right)
{}_{p_2}C_{l_2} \cdots {}_{p_{9}}C_{l_{9}}~.
\label{nnu-Z3-P}
\end{eqnarray}
The formulae for $l_L$, $(u_R)^c$, $(e_R)^c$ and $q_L$ are
obtained by replacing the summation of $(l_1, l_2)$ for $n_{\bar{d}}$
with $\{(3,1),(0,1)\}$, $\{(2,0),(1,2)\}$, $\{(0,2),(3,0)\}$
and $\{(1,1),(2,1)\}$. 

Here, we give some examples for representations and BCs to derive three families of SM fermions, in Table \ref{T7}. 
\begin{table}[htbp]
\caption{Examples for the three families of SM multiplets from $T^2/Z_3$.}
\label{T7}
\begin{center}
\begin{tabular}{|c|c|c|c|}  
\hline
$[N,k]$&$(p_1,p_2,p_3,p_4,p_5,p_6,p_7,p_8,p_9)$&$(\alpha_+,\beta_+)$&$(\alpha_-,\beta_-)$\\
\hline \hline
[11,4]&(3,2,0,0,1,2,3,0,0)&(0,1)&(0,1)\\
\hline
[12,3]&(3,2,0,1,1,0,1,2,2)&(1,0)&(0,1) \\
\hline
\end{tabular}
\end{center}
\end{table}

\subsection{$T^2/Z_4$}

For the representation matrices given by
\begin{eqnarray}
\hspace{-1.5cm}&~& Q_{0} = {\mbox{diag}}([+1]_{p_1},[+1]_{p_2},[+i]_{p_3},[+i]_{p_4},[-1]_{p_5},[-1]_{p_6},
[-i]_{p_7},[-i]_{p_8})~,
\nonumber \\
\hspace{-1.5cm}&~& P_{1} = {\mbox{diag}}([+1]_{p_1},[-1]_{p_2},[+1]_{p_3},[-1]_{p_4},[+1]_{p_5},[-1]_{p_6},
[+1]_{p_7},[-1]_{p_8})~,
\label{Z4-QP}
\end{eqnarray}
the following breakdown of $SU(N)$ gauge symmetry occurs
\begin{eqnarray}
SU(N) \to SU(p_1)\times SU(p_2) \times \cdots \times SU(p_8)\times U(1)^{7-n}~,
\label{Z4-GSB}
\end{eqnarray}
where $[\pm 1]_{p_i}$ and $[\pm i]_{p_i}$ represent 
$\pm 1$ and $\pm i$ for all $p_i$ elements. 

After the breakdown of $SU(N)$, $[N, k]_{\pm}$
is decomposed as
\begin{eqnarray}
[N, k]_{\pm} = \sum_{l_1 =0}^{k} \sum_{l_2 = 0}^{k-l_1} \cdots \sum_{l_7 = 0}^{k-l_1-\cdots -l_6}  
\left({}_{p_1}C_{l_1}, {}_{p_2}C_{l_2}, \cdots, {}_{p_8}C_{l_8}\right)_{\pm}~,
\label{Nk-Z4}
\end{eqnarray}
where $l_8=k-l_1- \cdots -l_7$.
The $\left({}_{p_1}C_{l_1}, {}_{p_2}C_{l_2}, \cdots, {}_{p_8}C_{l_8}\right)_{\pm}$ 
has the $Z_4$ and $Z_2$ elements
\begin{eqnarray}
&~& \mathcal{P}_{0 \pm} = i^{l_3+l_4} (-1)^{l_5+l_6} (-i)^{l_7+l_8} \eta^0_{k \pm}
= i^{l_1+l_2 + 2(l_3+l_4) + 3(l_5+l_6)} (-i)^k \eta^0_{k \pm}
\nonumber \\
&~& ~~~~~~~~ = i^{l_1+l_2 + 2(l_3+l_4) + 3(l_5+l_6) + \alpha_{\pm}}~, 
\label{Z40}\\
&~& \mathcal{P}_1 = (-1)^{l_2+l_4+l_6+l_8} \eta^1_{k \pm}
= (-1)^{l_1+l_3+l_5+l_7} (-1)^k \eta^1_{k \pm}
\nonumber \\
&~& ~~~~~~ = (-1)^{l_1+l_3+l_5+l_7 + \beta_{\pm}} ~,
\label{Z41}
\end{eqnarray}
where $\eta^0_{k \pm}$ takes a value $1$, $-1$, $i$ or $-i$,
and we parameterize the intrinsic $Z_M$ elements $(M=4, 2)$ as $(-i)^k \eta^0_{k \pm} = i^{\alpha_{\pm}}$ 
and $(-1)^k \eta^1_{k \pm}= (-1)^{\beta_{\pm}}$.

\subsubsection{Numbers of $SU(5)$ multiplets on $T^2/Z_4$}

After the breakdown of $SU(N) \to  SU(5) \times SU(p_2) \times \cdots \times SU(p_8) \times U(1)^{7-m}$, 
$[N, k]_{\pm}$ is decomposed as
\begin{eqnarray}
[N, k]_{\pm} = \sum_{l_1 =0}^{k} \sum_{l_2 = 0}^{k-l_1} \cdots \sum_{l_7 = 0}^{k-l_1-\cdots -l_6}  
\left({}_{5}C_{l_1}, {}_{p_2}C_{l_2}, \cdots, {}_{p_8}C_{l_8}\right)_{\pm}~.
\label{Nk(p1=5)Z4}
\end{eqnarray}

Using the assignment of $Z_4$ and $Z_2$ element (\ref{Z40}) and (\ref{Z41}),
we find that zero modes appear if the following relations are satisfied,
\begin{eqnarray}
&~& n^0_{l_1 L \pm} \equiv l_2+2(l_3+l_4)+3(l_5+l_6) = 4-l_1 - \alpha_{\pm}~~~(\mbox{mod}~4)~,~~
\nonumber \\
&~& n^1_{l_1 L \pm} \equiv l_3+l_5+l_7 = 2- l_1 - \beta_{\pm}~~~(\mbox{mod}~2)~.
\label{sum-Z4L}
\end{eqnarray}
The relation $n^a_{l_1 R \pm}=n^a_{l_1 L \pm} \mp 1$ (mod $4$) holds from (\ref{rhoPsiR}).

Then, we obtain the formulae (\ref{nbar5-ZM}) -- (\ref{n1-ZM}) with $n=8$, and
they are rewritten as
\begin{eqnarray}
\hspace{-1cm}&~& n_{\bar{5}} = \sum_{l_1 = 1, 4} \sum_{l_2=0}^{k-l_1} \cdots 
 \sum_{l_7=0}^{k-l_1-\cdots -l_6} (-1)^{l_1} 
\left(P^{(1,1)}_{+} - P^{(i,-1)}_{+} + P^{(1,1)}_{-} - P^{(-i,-1)}_{-}\right)
{}_{p_2}C_{l_2} \cdots {}_{p_{8}}C_{l_{8}}~, 
\label{nbar5-Z4-P}\\
\hspace{-1cm}&~& n_{10} = \sum_{l_1 = 2, 3} \sum_{l_2=0}^{k-l_1} \cdots 
\sum_{l_7=0}^{k-l_1-\cdots -l_6} (-1)^{l_1} 
\left(P^{(1,1)}_{+} - P^{(i,-1)}_{+} + P^{(1,1)}_{-} - P^{(-i,-1)}_{-}\right)
{}_{p_2}C_{l_2} \cdots {}_{p_{8}}C_{l_{8}}~, 
\label{n10-Z4-P}\\
\hspace{-1cm}&~& n_{1} = \sum_{l_1 = 0, 5} \sum_{l_2=0}^{k-l_1} \cdots 
\sum_{l_7=0}^{k-l_1-\cdots -l_6}
\left(P^{(1,1)}_{+} + P^{(i,-1)}_{+} + P^{(1,1)}_{-} + P^{(-i,-1)}_{-}\right)
{}_{p_2}C_{l_2} \cdots {}_{p_{8}}C_{l_{8}}~,
\label{n1-Z4-P}
\end{eqnarray}
where $P^{(\rho, \rho')}_{\pm}$ are projection operators that pick up the part relating
$(\mathcal{P}_{0 \pm}, \mathcal{P}_{1 \pm})=(\rho, \rho')$ and are written by
\begin{eqnarray}
P^{(\rho, \rho')}_{\pm} = \frac{1}{8}
\left(1+ \overline{\rho} \mathcal{P}_{0 \pm} + \overline{\rho}^2 \mathcal{P}_{0 \pm}^2
+ \overline{\rho}^3 \mathcal{P}_{0 \pm}^3\right)
\left(1+ \overline{\rho'} \mathcal{P}_{1 \pm}\right)~.
\label{P-Z4}
\end{eqnarray}

Here, we give some examples for representations and BCs to derive $n_{\bar{5}} = n_{10} = 3$, in Table \ref{T8}. 
\begin{table}[htbp]
\caption{Examples for the three families of $SU(5)$ from $T^2/Z_4$.}
\label{T8}
\begin{center}
\begin{tabular}{|c|c|c|c|}  
\hline
$[N,k]$&$(p_1,p_2,p_3,p_4,p_5,p_6,p_7,p_8)$&$(\alpha_+,\beta_+)$&$(\alpha_-,\beta_-)$\\
\hline \hline
[8,3]&(5,0,0,0,0,0,3,0)&(2,1)&(0,0)\\
\hline
[8,4]&(5,0,0,3,0,0,0,0)&(0,0)&(2,0) \\
\hline
[9,3]&(5,3,0,0,0,0,0,1)&(1,0)&(0,1) \\
\hline
[9,4]&(5,0,2,0,0,0,1,1)&(2,0)&(2,0) \\
\hline
[10,3]&(5,0,0,0,3,0,0,2)&(1,0)&(2,0) \\
\hline
[10,4]&(5,0,0,0,0,4,0,1)&(0,0)&(2,1) \\
\hline
[11,3]&(5,0,0,1,2,2,0,1)&(3,1)&(2,0) \\
\hline
[11,4]&(5,0,3,1,2,0,0,0)&(2.0)&(1,1) \\
\hline
[11,5]&(5,0,0,2,0,0,1,3)&(0,1)&(3,0) \\
\hline
[12,3]&(5,4,0,1,0,0,0,2)&(3,1)&(1,0) \\
\hline
[12,4]&(5,0,4,0,1,2,0,0)&(2,0)&(3,0) \\
\hline
[12,5]&(5,1,2,0,2,2,0,0)&(3,1)&(1,1) \\
\hline
[12,6]&(5,0,3,0,1,0,3,0)&(2,0)&(2,1) \\
\hline
\end{tabular}
\end{center}
\end{table}

\subsubsection{Numbers of the SM multiplets on $T^2/Z_4$}

After the breakdown of $SU(N) \to  SU(3) \times SU(2) \times SU(p_3) \times \cdots \times SU(p_8) \times U(1)^{7-m}$, 
$[N, k]_{\pm}$ is decomposed as
\begin{eqnarray}
[N, k]_{\pm} = \sum_{l_1 =0}^{k} \sum_{l_2 = 0}^{k-l_1} \cdots \sum_{l_7 = 0}^{k-l_1-\cdots -l_6}  
\left({}_{3}C_{l_1}, {}_{2}C_{l_2}, {}_{p_3}C_{l_3}, \cdots, {}_{p_8}C_{l_8}\right)_{\pm}~.
\label{Nk(p1=3)Z4}
\end{eqnarray}

Using the assignment of $Z_4$ and $Z_2$ element (\ref{Z40}) and (\ref{Z41}),
we find that zero modes appear if the following relations are satisfied,
\begin{eqnarray}
&~& n^0_{l_1 l_2 L \pm} \equiv 2(l_3+l_4)+3(l_5+l_6) = 4-l_1 -l_2- \alpha_{\pm}~~~(\mbox{mod}~4)~,~~
\nonumber \\
&~& n^1_{l_1 l_2 L \pm} \equiv l_3+l_5+l_7 = 2- l_1 - \beta_{\pm}~~~(\mbox{mod}~2)~.
\label{sum-Z4L-SM}
\end{eqnarray}
The relation $n^a_{l_1 l_2 R \pm}=n^a_{l_1 l_2 L \pm} \mp 1$ (mod $4$) holds from (\ref{rhoPsiR}).

Then, we obtain the formulae (\ref{nbar5-ZM}) -- (\ref{n1-ZM}) with $n=8$.
Using the projection operators (\ref{P-Z4}), 
the formulae for $(d_R)^c$ and $(\nu_R)^c$ are rewritten as
\begin{eqnarray}
\hspace{-1.3cm}&~& n_{\bar{d}} = \!\!\!\!\!\!\! \sum_{(l_1, l_2) = (2,2),(1,0)}
\sum_{l_2=0}^{k-l_1} \cdots \!\!\!\!\! \sum_{l_7=0}^{k-l_1-\cdots -l_6}
\!\!\!\!\!\!\!\!\!\!\! (-1)^{l_1+l_2} 
\left(P^{(1,1)}_{+} - P^{(i,-1)}_{+} + P^{(1,1)}_{-} - P^{(-i,-1)}_{-}\right)
{}_{p_2}C_{l_2} \cdots {}_{p_{8}}C_{l_{8}}~, 
\label{nd-Z4-P}\\
\hspace{-1.3cm}&~& n_{\bar{\nu}} = \!\!\!\!\!\!\! \sum_{(l_1, l_2) = (0,0),(3,2)} 
 \sum_{l_2=0}^{k-l_1} \cdots \!\!\!\!\! \sum_{l_7=0}^{k-l_1-\cdots -l_6}
\left(P^{(1,1)}_{+} + P^{(i,-1)}_{+} + P^{(1,1)}_{-} + P^{(-i,-1)}_{-}\right)
{}_{p_2}C_{l_2} \cdots {}_{p_{8}}C_{l_{8}}~.
\label{nnu-Z4-P}
\end{eqnarray}
The formulae for $l_L$, $(u_R)^c$, $(e_R)^c$ and $q_L$ are
obtained by replacing the summation of $(l_1, l_2)$ for $n_{\bar{d}}$
with $\{(3,1),(0,1)\}$, $\{(2,0),(1,2)\}$, $\{(0,2),(3,0)\}$
and $\{(1,1),(2,1)\}$. 

Here, we give some examples of representations and BCs to derive three families of SM fermions, in Table \ref{T9}. 
\begin{table}[htbp]
\caption{Examples for the three families of SM multiplets from $T^2/Z_4$.}
\label{T9}
\begin{center}
\begin{tabular}{|c|c|c|c|}  
\hline
$[N,k]$&$(p_1,p_2,p_3,p_4,p_5,p_6,p_7,p_8)$&$(\alpha_+,\beta_+)$&$(\alpha_-,\beta_-)$\\
\hline \hline
[9,3]&(3,2,1,0,0,0,2,1)&(0,1)&(0,0)\\
\hline
[11,3]&(3,2,1,1,0,4,0,0)&(1,0)&(1,1) \\
\hline
[11,4]&(3,2,0,0,3,1,1,1)&(0,1)&(0,0) \\
\hline
[12,4]&(3,2,1,0,2,1,3,0)&(0,1)&(0,0) \\
\hline
[12,6]&(3,2,1,2,0,0,0,4)&(0,1)&(1,1) \\
\hline
[13,4]&(3,2,1,2,2,2,0,1)&(0,1)&(0,0) \\
\hline
\end{tabular}
\end{center}
\end{table}

\subsection{$T^2/Z_6$}

For the representation matrices given by
\begin{eqnarray}
\hspace{-1.5cm}&~& \Xi_{0} 
= {\mbox{diag}}([+1]_{p_1},[+1]_{p_2},[\varphi]_{p_3},[\varphi]_{p_4},[\varphi^2]_{p_5},[\varphi^2]_{p_6},
\nonumber \\
\hspace{-1.5cm}&~& ~~~~~~~~~~~~~~~~~~~
[-1]_{p_7},[-1]_{p_8}, [-\varphi]_{p_9},[-\varphi]_{p_{10}}, [-\varphi^2]_{p_{11}},[-\varphi^2]_{p_{12}})~,
\nonumber \\
\hspace{-1.5cm}&~& P_{1} = {\mbox{diag}}([+1]_{p_1},[-1]_{p_2},[+1]_{p_3},[-1]_{p_4},[+1]_{p_5},[-1]_{p_6},
\nonumber \\
\hspace{-1.5cm}&~& ~~~~~~~~~~~~~~~~~~~
[+1]_{p_7},[-1]_{p_8}, [+1]_{p_9},[-1]_{p_{10}}, [+1]_{p_{11}},[-1]_{p_{12}})~,
\label{Z6-XiP}
\end{eqnarray}
the following breakdown of $SU(N)$ gauge symmetry occurs
\begin{eqnarray}
SU(N) \to SU(p_1)\times SU(p_2) \times \cdots \times SU(p_{12})\times U(1)^{11-m}~,
\label{Z6-GSB}
\end{eqnarray}
where $\varphi = e^{\pi i/3}$ and $[c]_{p_i}$ represents the number $c$ for all $p_i$ elements. 

After the breakdown of $SU(N)$, $[N, k]_{\pm}$,
is decomposed as
\begin{eqnarray}
[N, k]_{\pm} = \sum_{l_1 =0}^{k} \sum_{l_2 = 0}^{k-l_1} \cdots \sum_{l_{11} = 0}^{k-l_1-\cdots -l_{10}}  
\left({}_{p_1}C_{l_1}, {}_{p_2}C_{l_2}, \cdots, {}_{p_{12}}C_{l_{12}}\right)_{\pm}~,
\label{Nk-Z6}
\end{eqnarray}
where $l_{12}=k-l_1- \cdots -l_{11}$.
The $\left({}_{p_1}C_{l_1}, {}_{p_2}C_{l_2}, \cdots, {}_{p_{12}}C_{l_{12}}\right)_{\pm}$ 
has the $Z_6$ and $Z_2$ elements 
\begin{eqnarray}
&~& \mathcal{P}_0 = \varphi^{l_3+l_4} (\varphi^2)^{l_5+l_6} (-1)^{l_7+l_8}
(-\varphi)^{l_9+l_{10}} (-\varphi^2)^{l_{11}+l_{12}} \eta^0_{k \pm}
\nonumber \\
&~& ~~~~~~ = \varphi^{l_1+l_2+2(l_3+l_4)+3(l_5+l_6)+4(l_7+l_8)+5(l_9+l_{10})} \overline{\varphi}^k \eta^0_{k \pm}
\nonumber \\
&~& ~~~~~~ = \varphi^{l_1+l_2+2(l_3+l_4)+3(l_5+l_6)+4(l_7+l_8)+5(l_9+l_{10})+\alpha_{\pm}}~, 
\label{Z60}\\
&~& \mathcal{P}_1 = (-1)^{l_2+l_4+l_6+l_8+l_{10}+l_{12}} \eta^1_{k \pm}
= (-1)^{l_1+l_3+l_5+l_7+l_9+l_{11}} (-1)^k \eta^1_{k \pm}
\nonumber \\
&~& ~~~~~~ = (-1)^{l_1+l_3+l_5+l_7+l_9+l_{11}+\beta_{\pm}} ~,
\label{Z61}
\end{eqnarray}
where $\eta^0_{k \pm}$ takes a value $e^{n \pi i/3}$ $(n=0,1, \cdots, 5)$,
and we parameterize the intrinsic $Z_M$ elements $(M=6, 2)$ as 
$(e^{-\pi i/3})^k \eta^0_{k \pm} = (e^{\pi i/3})^{\alpha_{\pm}}$ and $(-1)^k \eta^1_{k \pm} = (-1)^{\beta_{\pm}}$.

\subsubsection{Numbers of $SU(5)$ multiplets on $T^2/Z_6$}

After the breakdown of $SU(N) \to  SU(5) \times SU(p_2) \times \cdots \times SU(p_{12}) \times U(1)^{11-m}$, 
$[N, k]_{\pm}$ is decomposed as
\begin{eqnarray}
[N, k]_{\pm} = \sum_{l_1 =0}^{k} \sum_{l_2 = 0}^{k-l_1} \cdots \sum_{l_{11} = 0}^{k-l_1-\cdots -l_{10}}  
\left({}_{5}C_{l_1}, {}_{p_2}C_{l_2}, \cdots, {}_{p_{12}}C_{l_{12}}\right)_{\pm}~.
\label{Nk(p1=5)Z6}
\end{eqnarray}

Using the assignment of $Z_6$ and $Z_2$ element (\ref{Z60}) and (\ref{Z61}),
we find that zero modes appear if the following relations are satisfied,
\begin{eqnarray}
\hspace{-1cm}&~& n^0_{l_1 L \pm} \equiv l_2+2(l_3+l_4)+3(l_5+l_6) +4(l_7+l_8)+5(l_9+l_{10})
= 6-l_1 - \alpha_{\pm}~~~(\mbox{mod}~6)~,~~
\nonumber \\
\hspace{-1cm}&~& n^1_{l_1 L \pm} \equiv l_3+l_5+l_7+l_9+l_{11} = 2- l_1 - \beta_{\pm}~~~(\mbox{mod}~2)~.
\label{sum-Z6L}
\end{eqnarray} 
The relation $n^a_{l_1 R \pm}=n^a_{l_1 L \pm} \mp 1$ (mod $6$) holds from (\ref{rhoPsiR}).

Then, we obtain the formulae (\ref{nbar5-ZM}) -- (\ref{n1-ZM}) with $n=12$, and
they are rewritten as
\begin{eqnarray}
\hspace{-1.2cm}&~& n_{\bar{5}} = \sum_{l_1 = 1, 4} \sum_{l_2=0}^{k-l_1} \cdots 
\!\!\! \sum_{l_{11}=0}^{k-l_1-\cdots -l_{10}}
\!\!\!\!\!\!\!\! (-1)^{l_1} \left(P^{(1,1)}_{+} - P^{(\varphi,-1)}_{+} 
+ P^{(1,1)}_{-} - P^{(\overline{\varphi},-1)}_{-}\right)
{}_{p_2}C_{l_2} \cdots {}_{p_{12}}C_{l_{12}}~, 
\label{nbar5-Z6-P}\\
\hspace{-1.2cm}&~& n_{10} = \sum_{l_1 = 2, 3} \sum_{l_2=0}^{k-l_1} \cdots 
\!\!\! \sum_{l_{11}=0}^{k-l_1-\cdots -l_{10}}
\!\!\!\!\!\!\!\! (-1)^{l_1} \left(P^{(1,1)}_{+} - P^{(\varphi,-1)}_{+} 
+ P^{(1,1)}_{-} - P^{(\overline{\varphi},-1)}_{-}\right)
{}_{p_2}C_{l_2} \cdots {}_{p_{12}}C_{l_{12}}~, 
\label{n10-Z6-P}\\
\hspace{-1.2cm}&~& n_{1} = \sum_{l_1 = 0, 5} \sum_{l_2=0}^{k-l_1} \cdots 
\!\!\! \sum_{l_{11}=0}^{k-l_1-\cdots -l_{10}}
\!\!\!\!\!\!\!\! \left(P^{(1,1)}_{+} + P^{(\varphi,-1)}_{+} 
+ P^{(1,1)}_{-} + P^{(\overline{\varphi},-1)}_{-}\right)
{}_{p_2}C_{l_2} \cdots {}_{p_{12}}C_{l_{12}}~,
\label{n1-Z6-P}
\end{eqnarray}
where $P^{(\rho, \rho')}_{\pm}$ are projection operators that pick up the part relating
$(\mathcal{P}_{0 \pm}, \mathcal{P}_{1 \pm})=(\rho, \rho')$ and are written by
\begin{eqnarray}
P^{(\rho, \rho')}_{\pm} = \frac{1}{12}
\left(1+ \overline{\rho} \mathcal{P}_{0 \pm} + \overline{\rho}^2 \mathcal{P}_{0 \pm}^2
+ \overline{\rho}^3 \mathcal{P}_{0 \pm}^3 + \overline{\rho}^4 \mathcal{P}_{0 \pm}^4
+ \overline{\rho}^5 \mathcal{P}_{0 \pm}^5\right)
\left(1+ \overline{\rho'} \mathcal{P}_{1 \pm}\right)~.
\label{P-Z6}
\end{eqnarray}

Here, we give some examples for representations and BCs to derive $n_{\bar{5}} = n_{10} = 3$, in Table \ref{T10}. 
\begin{table}[htbp]
\caption{Examples for the three families of $SU(5)$ from $T^2/Z_6$.}
\label{T10}
\begin{center}
\begin{tabular}{|c|c|c|c|}  
\hline
$[N,k]$&$(p_1,p_2,p_3,p_4,p_5,p_6,p_7,p_8,p_9,p_{10},p_{11},p_{12})$&$(\alpha_+,\beta_+)$&$(\alpha_-,\beta_-)$\\
\hline \hline
[8,3]&(5,0,0,3,0,0,0,0,0,0,0,0)&(0,1)&(2,0)\\
\hline
[8,4]&(5,0,0,1,0,0,0,2,0,0,0,0)&(0,0)&(2,0) \\
\hline
[9,3]&(5,0,0,0,0,0,3,0,0,0,0,1)&(0,1)&(5,0) \\
\hline
[9,4]&(5,2,0,1,0,0,1,0,0,0,0,0)&(2,0)&(2,0) \\
\hline
[10,3]&(5,0,0,1,1,0,0,0,0,0,3,0)&(0,1)&(4,1) \\
\hline
[10,4]&(5,0,1,0,1,1,0,0,0,1,1,0)&(5,0)&(2,0) \\
\hline
[10,5]&(5,0,0,0,0,0,1,2,0,2,0,0)&(4,1)&(1,0) \\
\hline
[11,3]&(5,0,0,1,0,0,0,0,0,1,4,0)&(3,1)&(4,1) \\
\hline
[11,4]&(5,0,0,0,0,2,0,0,2,1,0,1)&(5,0)&(2,0) \\
\hline
[11,5]&(5,3,0,0,0,0,0,0,0,0,3,0)&(1,1)&(1,1) \\
\hline
[12,3]&(5,3,0,1,0,0,0,0,0,0,0,3)&(0,1)&(3,0) \\
\hline
[12,4]&(5,0,0,0,0,0,0,1,0,4,1,1)&(5,0)&(2,0) \\
\hline
[12,5]&(5,0,0,0,0,0,2,1,2,1,1,0)&(1,1)&(1,1) \\
\hline
[12,6]&(5,0,0,0,0,3,1,1,2,0,0,0)&(3,0)&(0,0) \\
\hline
\end{tabular}
\end{center}
\end{table}

\subsubsection{Numbers of the SM multiplets on $T^2/Z_6$}

After the breakdown of $SU(N) \to  SU(3) \times SU(2) \times SU(p_2) \times \cdots \times SU(p_{12}) \times U(1)^{11-m}$, 
$[N, k]_{\pm}$ is decomposed as
\begin{eqnarray}
[N, k]_{\pm} = \sum_{l_1 =0}^{k} \sum_{l_2 = 0}^{k-l_1} \cdots \sum_{l_{11} = 0}^{k-l_1-\cdots -l_{10}}  
\left({}_{3}C_{l_1}, {}_{2}C_{l_2}, {}_{p_3}C_{l_3}, \cdots, {}_{p_{12}}C_{l_{12}}\right)_{\pm}~.
\label{Nk(p1=3)Z6}
\end{eqnarray}

Using the assignment of $Z_6$ and $Z_2$ element (\ref{Z60}) and (\ref{Z61}),
we find that zero modes appear if the following relations are satisfied,
\begin{eqnarray}
\hspace{-1cm}&~& n^0_{l_1 l_2 L \pm} \equiv 2(l_3+l_4)+3(l_5+l_6) +4(l_7+l_8)+5(l_9+l_{10})
= 6-l_1 -l_2- \alpha_{\pm}~~~(\mbox{mod}~6)~,~~
\nonumber \\
\hspace{-1cm}&~& n^1_{l_1 l_2 L \pm} \equiv l_3+l_5+l_7+l_9+l_{11} = 2- l_1 - \beta_{\pm}~~~(\mbox{mod}~2)~.
\label{sum-Z6L-SM}
\end{eqnarray}
The relation $n^a_{l_1 l_2 R \pm}=n^a_{l_1 l_2 L \pm} \mp 1$ (mod $6$) holds from (\ref{rhoPsiR}).

Then, we obtain the formulae (\ref{nbar5-ZM}) -- (\ref{n1-ZM}) with $n=12$.
Using the projection operators (\ref{P-Z6}), 
the formulae for $(d_R)^c$ and $(\nu_R)^c$ are rewritten as
\begin{eqnarray}
\hspace{-1.6cm}&~& n_{\bar{d}} = \sum_{(l_1, l_2) = (2,2),(1,0)}
\sum_{l_2=0}^{k-l_1} \cdots \sum_{l_{11}=0}^{k-l_1-\cdots -l_{12}}
(-1)^{l_1+l_2} \left(P^{(1,1)}_{+} - P^{(\varphi,-1)}_{+} 
+ P^{(1,1)}_{-} - P^{(\overline{\varphi},-1)}_{-}\right)
\nonumber \\
\hspace{-1.6cm}&~& ~~~~~~~~~~~~~~~~~~~~~~~~~~~~~~~~~~~~~~~~~~~~~~~~~~~~~~~~~~~~~~~~~~~~~~~~~~ \times 
{}_{p_2}C_{l_2} \cdots {}_{p_{12}}C_{l_{12}}~, 
\label{nd-Z6-P}\\
\hspace{-1.6cm}&~& n_{\bar{\nu}} = \sum_{(l_1, l_2) = (0,0),(3,2)} 
 \sum_{l_2=0}^{k-l_1} \cdots \sum_{l_{11}=0}^{k-l_1-\cdots -l_{10}}
\left(P^{(1,1)}_{+} + P^{(\varphi,-1)}_{+} 
+ P^{(1,1)}_{-} + P^{(\overline{\varphi},-1)}_{-}\right)
{}_{p_2}C_{l_2} \cdots {}_{p_{12}}C_{l_{12}}~.
\label{nnu-Z6-P}
\end{eqnarray}
The formulae for $l_L$, $(u_R)^c$, $(e_R)^c$ and $q_L$ are
obtained by replacing the summation of $(l_1, l_2)$ for $n_{\bar{d}}$
with $\{(3,1),(0,1)\}$, $\{(2,0),(1,2)\}$, $\{(0,2),(3,0)\}$
and $\{(1,1),(2,1)\}$.

Here, we give some examples for representations and BCs to derive three families of SM fermions, in Table \ref{T11}. 
\begin{table}[htbp]
\caption{Examples for the three families of SM multiplets from $T^2/Z_6$.}
\label{T11}
\begin{center}
\begin{tabular}{|c|c|c|c|}  
\hline
$[N,k]$&$(p_1,p_2,p_3,p_4,p_5,p_6,p_7,p_8,p_9,p_{10},p_{11},p_{12})$&$(\alpha_+,\beta_+)$&$(\alpha_-,\beta_-)$\\
\hline \hline
[9,3]&(3,2,0,1,0,0,0,0,0,0,1,2)&(0,0)&(0,1)\\
\hline
[9,4]&(3,2,0,0,0,1,0,0,1,2,0,0)&(1,1)&(1,0) \\
\hline
[10,3]&(3,2,0,0,3,0,0,0,0,0,1,1)&(1,0)&(1,1) \\
\hline
[10,4]&(3,2,0,1,1,2,0,0,0,0,1,0)&(0,1)&(0,0) \\
\hline
[11,3]&(3,2,1,1,1,0,0,0,0,1,1,1)&(0,1)&(0,0) \\
\hline
[11,4]&(3,2,0,1,0,2,0,0,0,3,0,0)&(0,1)&(1,0) \\
\hline
[11,5]&(3,2,0,0,1,0,4,0,1,0,0,0)&(0,1)&(0,0) \\
\hline
[12,3]&(3,2,0,1,3,1,0,1,0,0,0,1)&(1,0)&(1,1) \\
\hline
[12,4]&(3,2,0,0,0,1,1,2,0,2,1,0)&(1,1)&(1,0) \\
\hline
[12,5]&(3,2,1,1,0,3,1,1,0,0,0,0)&(1,0)&(1,1) \\
\hline
[12,6]&(3,2,0,0,0,1,0,0,3,0,0,3)&(1,1)&(1,1) \\
\hline
[13,3]&(3,2,1,0,0,0,0,3,2,0,0,2)&(0,0)&(0,1) \\
\hline
[13,4]&(3,2,2,0,1,1,1,1,0,0,1,1)&(1,0)&(1,1) \\
\hline
[13,5]&(3,2,1,0,0,4,0,0,0,3,0,0)&(1,1)&(1,0) \\
\hline
[13,6]&(3,2,1,0,0,0,0,2,4,0,0,1)&(0,0)&(0,1) \\
\hline
\end{tabular}
\end{center}
\end{table}

\section{Conclusions}

We have studied the possibility of family unification on the basis of $SU(N)$ gauge theory
on the 6-dimensional space-time, $M^4\times T^2/Z_N$.
We have obtained enormous numbers of models with three families of $SU(5)$ matter multiplets 
and those with three families of the SM multiplets,
from a single massless Dirac fermion with a higher-dimensional representation of $SU(N)$, 
after the orbifold breaking.
Total numbers of models with the three families of $SU(5)$ multiplets and the SM multiplets
are summarized in Table \ref{T12} and \ref{T13}, respectively.
Our results can give a starting point for the construction toward a more realistic model,
because three families of chiral fermions in the SM standard model contain in our models.

Now, the following open questions should be tackled as a future work.

The unwanted matter degrees of freedom can be successfully made massive thanks to the orbifolding. 
However, some extra gauge fields remain massless, 
even after the symmetry breaking due to the Hosotani mechanism~\cite{Hosotani1,Hosotani2}. 
In most cases, this kind of non-abelian gauge subgroup plays the role of family symmetry.
These massless degrees of freedom must be made massive by further breaking of the family symmetry.
Extra scalar fields can play a role of Higgs fields 
for the breakdown of extra gauge symmetries including non-abelian gauge symmetries.
As a result, extra massless fields including the family gauge bosons can be massive.

In general, there appear $D$-term contributions to scalar masses in supersymmetric models 
after the breakdown of such extra gauge symmetries 
and the $D$-term contributions lift the mass degeneracy.~\cite{D,H&K,KM&Y1,KM&Y2,K&T}.
The mass degeneracy for each squark and slepton species in the first two families is favorable for suppressing
flavor-changing neutral current (FCNC) processes.
The dangerous FCNC processes can be avoided if the sfermion masses in the first two families are rather large 
or the fermion and its superpartner mass matrices are aligned.
The requirement of degenerate masses would yield a constraint on the $D$-term condensations
and/or SUSY breaking mechanism unless other mechanisms work.
If we consider the Scherk-Schwarz mechanism~\cite{S&S,S&S2} for $N=1$ SUSY breaking, 
the $D$-term condensations can vanish for the gauge symmetries broken at the orbifold breaking scale,
because of a universal structure of the soft SUSY breaking parameters.
The $D$-term contributions have been studied in the framework of $SU(N)$ orbifold GUTs~\cite{KK,KK&M}.

Can the gauge coupling unification successfully achieved?
If the particle contents in the minimal supersymmetric standard model only remain in the low-energy spectrum 
around and below the TeV scale and a big desert exists after the breakdown of extra gauge symmetries, 
an ordinary grand unification scenario can be realized up to the threshold corrections due to the Kaluza-Klein modes 
and the brane contributions from non-unified gauge kinetic terms.

Another problem is whether or not the realistic fermion mass spectrum and the generation mixings are successfully achieved.
Fermion mass hierarchy and generation mixings can also occur through the Froggatt-Nielsen mechanism~\cite{FN}
on the breakdown of extra gauge symmetries and the suppression of brane-localized Yukawa coupling constants 
among brane weak Higgs doublets and bulk matters with the volume suppression factor~\cite{Y}.

It would be interesting to reconsider or reconstruct our models in the framework of string theory.
Various 4-dimensional string models including three families have been constructed from several methods,
see e.g.~\cite{I&U} and references therein for useful articles.\footnote{
See also Ref.~\cite{M&Y} and references therein for recent works.
}

Furthermore, it would be interesting to study cosmological implications of the class of models presented in this paper, 
see e.g.~\cite{Khlopov:1999rs} and references therein for useful articles toward this direction.

\section*{Acknowledgements}
This work was supported in part by scientific grants from the Ministry of Education, Culture,
Sports, Science and Technology under Grant Nos.~22540272 and 21244036 (Y.K.).

\end{document}